\documentstyle[11pt,psfig]{article}

\normalbaselines
\font\tbf = cmbx12
\pagebreak
\input{tcilatex}
\begin{document}

\tableofcontents
\listoffigures

\begin{titlepage}
\indent
\vskip 0.1cm
\centerline{\tbf THE SLOPE EQUATIONS: A UNIVERSAL RELATIONSHIP BETWEEN}
\centerline{}
\centerline{\tbf LOCAL STRUCTURE AND GLASS TRANSITION TEMPERATURE}
\vskip 0.4cm
\centerline{} 
\vskip 0.5cm
\centerline{Matthieu Micoulaut}
\vskip 1cm
\centerline{Laboratoire GCR - CNRS URA 769}
\centerline{Universit\'e Pierre et Marie Curie, Tour 22, Boite 142}
\centerline{4, Place Jussieu, 75005 Paris, France}
\vskip 0.4cm
\indent
{\bf Abstract:}\par
In this article, we present a universal relationship between the glass
transition temperature $T_g$ and the local glass structure. The derivation
of the simplest expression of this relationship and some 
comparisons with experimental $T_g$ values have already been reported 
in a recent letter \cite{l1}. We give here the analytical expression of the
parameter $\beta$ of the Gibbs-Di Marzio equation and also new
experimental probes for the validity of the relationship, especially in
low modified binary glasses. The influence of medium range order is presented 
and the unusual behavior of $T_g$ in binary $B_2S_3$ and $P_2S_5$ systems explained
by the presence of modifier-rich clusters (denoted by $B-B$ {\em doublets}).
\par
{\bf Pacs:} 61.20N-81.20P
\end{titlepage}

\section{Introduction}

It is well-known that the formation of glasses requires cooling to a
sufficiently low temperature -below the glass transition- without the
occurrence of detectable crystallization. In treating this phenomenon, it is
has been suggested by different authors that specific structural features or
physical properties will result in glasses being formed \cite{f1}-\cite{f6}. Various
models have been proposed in order to describe this transition, which can
appear superficially to be a second-order thermodynamic phase transition.
These models involve generally factors (thermodynamic, structural, kinetic)
which are viewed as decisive in the formation of glasses.

The best known structural model of glass formation and glass formation
ability is that due to Zachariasen \cite{f7} and others \cite{f8,f9} who
proposed a classification of oxide materials in terms of glass-formers (e.g. 
$SiO_2$), modifiers (e.g. $Li_2O$) and intermediates (e.g. $PbO$). This led
to the random network concept \cite{f10} which received support from x-ray
diffraction studies of a variety of glasses, although these studies did not
establish the model as unique representation of structure \cite{f11}-\cite
{f13}. In these models, the relationship with the glass transition 
temperature is lacking.

Importance of thermodynamic factors in glass formation has been pointed out
by Gibbs and Di Marzio and by Adam and Gibbs \cite{f14,f15}. These authors
suggested that the glassy state is thus defined in terms of thermodynamic
variables (temperature, volume,...) and related ones (bulk compressibility,
heat capacity,...), but it is not necessarily implied that the glassy state
is one of even metastable equilibrium (with reference to a possible
crystalline phase). It can be stated that a glass-forming material has
equilibrium properties, even if it may be difficult to realize. The theory
developed by Adam and Gibbs on this basis, was able to predict a
second-order phase transition and also a quantitative relationship between
glass transition temperature and the cross-linking density in some linear
molecular chains.

Nevertheless, when the glass transition temperature is measured under
standard conditions (for example calorimetrically at a fixed heating rate),
an important question concerns the relationship of $T_g$ with some other
physical and measurable factors. Various proposals have been made in the
past which suggested for example that $T_g$ scales with the melting
temperature (the '{\em two-third rule}` proposed by Kauzmann \cite{f16}), 
the boiling
temperature, the Debye temperature of the phonon spectrum, etc. \cite{f17}.
Besides the influence of these thermodynamic factors, attention has been
devoted to structural factors, in particular to the valence of the involved
atoms in the glass-forming material. Tanaka \cite{f18} has given an
empirical relationship between $T_g$ and $Z$, the average coordination
number per atom: $ln\ T_g\simeq 1.6\ Z+2.3$. This proposal agrees for
various glass-forming systems including chalcogenide materials and organic
polymeric materials. However, the relationship between structural factors
and $T_g$ becomes more complicated if the composition of the glass-forming
alloy is changing. For example, in $Ge_xSe_{1-x}$ 
network glasses, the glass transition temperature is not varying monotically
with $x$, and exhibit even some characteristic behavior (maximum at $x=0.33$ 
\cite{f19}, which corresponds to the stoichiometric composition $GeSe_2$,
anomaly at $x=0.2$, where the average coordination number $<r>$ is equal to
$2.4$).
Obviously, $T_g$ is sensitive to the chemistry involved, and a maximum in $%
T_g$ at the stoichiometric composition may result from the formation of a
chemically ordered network in which only the $Ge-Se$ bonds are present. In
other systems, a general accepted rule states that $T_g$ is increasing when
the connectivity of the network is increasing, and {\em vice-versa}. Besides
these rather qualitative relationships between structure and glass
transition temperature, there exists a firm rule for predicting $T_g$ in
particular glassy materials, namely chalcogenide network glasses. The
equation relating the glass transition temperature to some structural factor
is that due to Gibbs and Di Marzio \cite{f14} and Varshneya and co-workers 
have shown recently that a close equation was particularly well adapted for predicting $%
T_g$ in multicomponent chalcogenide glass systems \cite{f20}. Indeed, one can consider the
chalcogenide glass system as a network of chains (e.g. selenium atoms) in
which cross-linking units (such as germanium atoms) are inserted. The
increase of $T_g$ is produced by the growing presence of these cross-linking
agents, which can be roughly explained by the grwoth in network connectivity. 
In the former version, Gibbs and Di Marzio applied successfully their 
equation in ordre to explain the $T_g$ data in polymers \cite{f21}. An 
adapted theory constructed by Di Marzio 
\cite{f22} has shown that for glass systems with some chain stiffness, the
glass transition temperature versus cross-linking density $X$ could be
expressed as: 
\begin{equation}  \label{11}
T_g\ =\ {\frac {T_0}{1-\kappa X}}
\end{equation}
where $T_0$ is the glass transition temperature of the initial polymeric
chain and $\kappa$ a constant.

In this article, we shall present several relationships between the glass
transition temperature and the local structure in glass-forming materials,
by using an agglomeration model, created by R.Kerner and D.M.Dos Santos \cite
{f23} and applied with success to various glass-forming systems \cite{f24}.
The physical insight of the model can be found in references \cite{f25} and 
\cite{f26}. Different situations will be reviewed in this article and
previous results, reported in \cite{l1}, will be inserted for completeness.
In section 2, we present the construction with star-like entities and obtain
the first slope equation for corner-sharing structures (single bonded
network, i.e. absence of dimers).
Comparison with experimental data is presented in section 3 for chalcogenide
network glasses and for binary oxide glasses. The relationship with the Gibbs-Di
Marzio equation is also given in section 3. Section 4 is devoted to other
structural contributions, such as the edge-sharing tendency between two
local structures (two-membered rings, or dimers), which modifies slightly
the slope equation, and to the influence of particular bonds, leading to a
second set of slope equations. These equations can describe the unusual
behavior of $T_g$ in $B_2S_3$ and $P_2S_5$ based glasses. Finally, section 5 
summarizes the most important results of the paper.

\section{Construction with star-like entities}

\subsection{A very simple structural consideration}

For the reader's convenience, we shall first present the simplest possible
construction of the agglomeration model, with two star-like entities. The
vocabulary introduced in this section will be illustrated, whenever
possible, by two archetypal glass systems which are binary $SiO_2$ and $%
Ge_xSe_{1-x}$ glasses.

The most simplest way of describing short range order (SRO) in glasses can
be made on the basis of star-like entities (we shall also call {\em singlets}
or local configurations \cite{f26}). These local configurations share a
central atom and they have clear, unambigous experimental evidence and a
well-defined coordination number (fig. 1). The nature of the coordination
can eventually be revealed by X-ray or inelastic neutron diffraction
techniques \cite{r2}-\cite{r3} which exhibit sharp and characteristic peaks
at the corresponding bond length and give information about the number of
nearest possible neighbors of a central atom. Typical examples are: the
tetrahedron which is the lowest possible SRO structure in glasses such as $%
SiO_{2}$ or $SiSe_{2}$ \cite{r3} and their binary compounds (fig. 1), or a
four-valenced germanium atom in $Ge_{x}Se_{1-x}$ network glasses \cite{r5}.
Starting from a structure with a single type of {\em singlet}, called `` 
{\em regular} '' local configuration, one can modify the structure by adding
a second kind of star-like entity, denoted by `` {\em altered }'' local
configuration.
\par
The probability of finding a `` {\em regular} '' local configuration $A$
with coordination number $m$ can be denoted by $p$ (e.g. $SiO_{4/2}$
tetrahedra in $SiO_2-Li_2O$ systems and of course $m=4$), whereas the one
related to an `` {\em altered }'' atom $B$ with coordination number $%
m^{\prime}$ can be denoted by $1-p$. The coordination number $m^{\prime}$
can correspond either to the valence of the atom B, or to the number of
covalent bonds which connect the local configuration to the rest of the
glass network. In IV-VI based glasses, the number of coordination can
be obtained by considering $NMR$ spectroscopy patterns \cite{r7,r8}. For
example, when adding the modifier $Li_2O$ in $SiO_2$, the number of covalent 
$Si-O-Si$ bonds decreases, because of the creation of so-called "{\em %
non-bridging atoms}" (NBO) due to the high ionicity of lithium (ionic $Li^\oplus
O^\ominus$ bonds). The local configuration B should be the $%
SiO_{4/2}^\ominus Li^\oplus$ tetrahedron \cite{r7}, and obviously $%
m^{\prime}=3$ (figure 1). The creation of such a new local structure can be
proposed on the basis of $^{29}Si\ MAS\ NMR$ experiments. The peaks
attributed to the $SiO_{4/2}$ tetrahedra in $v-SiO_2$ (assigned to $Q^{(4)}$
structures in $NMR$ notation where "4" stands for the number of $Si-O-Si$
bridges) are slightly shifted with addition of $Li_2O$, and a typical 
chemical shift occurs due to the presence of $Q^{(3)}$ species (the $SiO_{4/2}^{\ominus}$
tetrahedron \cite{r7}) This chemical shift is compared to the one obtained 
from typical crystalline compounds (here the disilicate $Li_4Si_4O_{10}$)
and identified. Finally, the local structure in binary $SiO_2-Li_2O$
systems can be described in terms of $Q^{(k)}$ functions over the whole
concentration range \cite{r9} (where $4-k$ is the number of NBO's on a
tetrahedron). With these examples, it is easy to see that the local
configurations $A$ and $B$ can describe very well the structural change
induced by the addition of $Li_2O$ in silica based glasses, at least for
small concentrations of $Li_2O$. In the network glasses $A_xB_{1-x}$; the
number of coordination can be given by the $8-N$ rule (where $N$ is the
number of outer shell electrons)\cite{r10}.

We shall show how the glass transition temperature changes as a function of $%
m$ and $m^{\prime}$.

The aim of the construction is to evaluate the time dependence of the
fluctuations of the local configuration probability, and derive an equation
which imposes the minimization of local fluctuations \cite{f23}. One can
reasonabely assume that these fluctuations remain important as long as the
system is in the liquid or the supercooled state, where the low viscosity
allows still the (A,B) configuration interchange by movement, bond
destruction and creation or cation switching (in case of the presence of
alkali modifier). Therefore, this should produce
a variation of the local probability with respect to time and temperature.
In a high viscosity state, one should expect a vanishing of the
fluctuations, i.e. when the local probabilities reach a stationary value (no
more fluctuations of $A$ and $B$). This can be identified with a stable
(crystal) or a meta-stable solid (glass). We do not consider here relaxation
processes taking place just before or at $T_g$, as we are looking at a
stationnary r\'egime for $p$, and $t\rightarrow \infty$.
When $p$ is very small
(corresponding to a system with high proportion of $A$ configurations),
there are only two possible elementary processes of single bond formation
(fig. 2), i.e. $A-A$ and $A-B$, the second {\em doublet} being identical to $%
B-A$.
\par
The probabilities of these {\em doublets} may be proportional to the
products of the probabilities of {\em singlets}, a Boltzmann factor which
takes into account the energy of creation of the respective bond formation
(i.e. $E_{1}$ for $A-A$ and $E_{2}$ for $A-B$) and a statistical factor
which may be regarded as the degeneracy of the corresponding stored energy,
because there are several equivalent ways to join together two coordinated
local configurations \cite{f26}. In the single bond formation, the
statistical factor is simply the product of the corresponding coordination
numbers (see fig. 2). The probability of finding the constructed {\em %
doublets} is then: 
\begin{equation}
p_{AA1}\ =\ {\frac{m^{2}}{{\cal Z}_{1}}}(1-p)^{2}e^{-\beta E_{1}}  \label{21}
\end{equation}
\begin{equation}
p_{AB1}\ =\ {\frac{2mm^{\prime }}{{\cal Z}_{1}}}p(1-p)e^{-\beta E_{2}}
\label{22}
\end{equation}
where $\beta $ stands for $1/k_{B}T$ with $k_{B}$ the Boltzmann constant and 
${{\cal Z}_{1}}$ is the normalizing factor given by: 
\begin{equation}
{{\cal Z}_{1}}\ =\ m^{2}(1-p)^{2}e^{-\beta E_{1}}+2mm^{\prime
}p(1-p)e^{-\beta E_{2}}  \label{23}
\end{equation}
We have not considered here the possibility of creation of a bond
with two ''{\em altered} `` configurations $B$. We shall indeed first focus
our interest either on glass systems with a very low amount of local
configurations $B$ or on systems which do not possess the ability in
creating these bonds. These situations are observable in glasses with a
small concentration of modifier, e.g. $xLi_{2}O-(1-x)SiO_{2}$ systems with $%
x<0.2$ [existence of $Q^{(4)}$ and $Q^{(3)}$ species only \cite{r7}] or $%
Ge_{x}Se_{1-x} $ glasses with $x<0.33$ [existence of $Se-Se$ and $Ge-Se$
bonds only \cite{f19}]. The influence of $B-B$ bonds will be presented in a
forthcoming section. We have also excluded the possibility of a simultaneous
creation of two bonds $A-A$ or $A-B$ which leads to the formation of
two-membered rings (dimers); such possibility does exist in binary
chalcogenide glasses (e.g. $SiS_{2}$ based glasses), and will be also taken
into account below. Anyhow, the analysis remains exactly the same. The new 
{\em doublets} may create a local fluctuation in the statistics, which can
be evaluated as: 
\begin{equation}
1-p^{(1)}\ =\ {\frac{1}{2}}(2p_{AA1}+p_{AB1})  \label{24}
\end{equation}
\indent
If one denotes the average time needed to form a new bond by $\tau $, we can
evaluate the time derivative of $p$ due to the above fluctuation as 
\begin{equation}
{\frac{dp}{dt}}\ =\ {\frac{1}{\tau }}\biggl[p^{(1)}-p\biggr] \label{25}
\end{equation}
We have neglected the dependence of the cooling rate $q=dT/dt$ in the equation
(\ref{25}) because network chalcogenide glasses $%
A_{x}B_{1-x}$ or binary oxides such as $B_{2}O_{3}$ or $SiO_{2}$ based
glasses form very easily, and have critical cooling rates of the order of $%
10^{-4}K\,.s^{-1}$ \cite{r13}-\cite{r15}. For these systems, it seems
reasonable to neglect an additional cooling term.

When the local fluctuations are vanishing at $T_g$ and $t\rightarrow \infty$, 
the above expression
should be $0$, which amounts to finding the {\it stationary} or {\it singular%
} solutions of the differential equation (\ref{25}) This leads to: 
\begin{equation}  \label{28}
p (1 - p) \biggl[m(1-p)(m^{\prime}e^{-E_2/k_BT_g}- me^{-E_1/k_BT_g}) -
pmm^{\prime}e^{-E_2/k_BT_g})\biggr] = 0
\end{equation}
There are always two singular solutions at the points $p = 0$ and $p = 1$;
but there can exist also a third solution, given by the following
expression: 
\begin{equation}  \label{29}
p_{am} = \frac{m m^{\prime}e^{-E_2/k_BT_g} - m^2 e^{-E_1/k_BT_g}} {%
2mm^{\prime}e^{-E_2/k_BT_g} - m^2e^{-E_1/k_BT_g}}
\end{equation}
The solution $p_{am}$ represents a probability, it is therefore physically
acceptable only if $0 \leq p_{am} \leq 1$. Moreover, we want it to be an 
{\it attractive point}. Close to $p_{am}(T_g)$, every fluctuation $\xi$ should
indeed vanish for $t\rightarrow \infty$, hence the right-hand side of the
linearized equation (\ref{25}) in the vicinity of $p_{am}$ should be
negative. This is possible only if: 
\begin{equation}  \label{210}
\frac{m^{\prime}}{m} > e^{\frac{E_2 - E_1}{kT}}
\end{equation}
\indent
When condition (\ref{210}) is not satisfied,
the stable ({\it attractive}) solution is found at $p = 0$, which means that
at the microscopic level the agglomeration tends to separate the two kinds
of configurations, $A$ and $B$.

Note that due to the homogeneity of the expression (\ref{29}), only one
energy difference is essential here: $\alpha = (E_2 - E_1)/kT$ and that the
absence of a $B-B$ {\em doublet} yields automatically a repulsive solution
for $p=1$.

The phase portraits of the differential equation (\ref{25}) are shown in fig. 3.

\subsection{The first slope equation}

In order to give a more realistic meaning to the solution (\ref{29}), we
have to relate the probability $p_{am}$ of finding an '' {\em altered} ``
local configuration $B$, with the modifier concentration $x$. In case of
network glasses $A_xB_{1-x}$ (e.g. $Ge_xSe_{1-x}$), the identification is
obvious: $p_{am}=x$. For binary glass systems $(1-x)A_rX_s+xM_qX$
(or $A_rX_s+RM_qX$ following the notation), we have to
recall a charge conservation law \cite{f26,r19}. The concentration $x$ of $%
M^\oplus$ cations must be equal to the anionic contribution located on each
particular local configuration, expressed in terms of $p_{am}$: 
\begin{equation}  \label{211}
{\frac {1}{r}}x\ =\ {\frac {1}{q}}(1-x)\biggl[n_Bp_{am}+n_A(1-p_{am})\biggr]
\end{equation}
where $n_B$ and $n_A$ are the anionic contributions of the $B$ and $A$
configurations. For example, in low modified $(1-x)SiO_2-xLi_2O$ systems ($%
x<0.33$), equation (\ref{211}) becomes: 
\begin{equation}  \label{212}
R\ =\ {\frac {x}{1-x}}\ =\ {\frac {1}{2}}p_{am}
\end{equation}
where $p_{am}$ is the probability of finding $Q^{(3)}$ species and $R$ the
reduced concentration \cite{r19}. Since $x$ and $p_{am}$ are now related, there should
be various glass formers displaying the tendency towards the solution $%
p_{am} $ in a wide range of modifier concentration, also when it tends to
zero ($x \rightarrow 0$, i.e. $p_{am}\rightarrow 0$). In this limit, we can
obtain from (\ref{29}) a condition concerning the energy difference $E_2 -
E_1:$ 
\begin{equation}  \label{213}
E_2 - E_1 = k\,T_0\,ln\biggl[\frac{m^{\prime}}{m}\biggr]
\end{equation}
where $T_0$ is the glass transition temperature at $x = 0$, i.e. in the
limit when the modifier concentration goes to zero, corresponding to a 
pure A glass ($v-Se$ and $T_0\simeq 316\ K$) or a pure network former 
($SiO_2$ and $T_0\simeq 1450\ K$).

This equation exhibits a relation between the {\it statistical} and {\it %
energetic} factors that are crucial for the glass forming tendency to
appear. It tells us that in good binary glass formers whenever $m^{\prime}>m$
(and $ln(\frac{m^{\prime}}{m}) > 0$), one should expect $E_2 > E_1$, and
{\em vice versa}, in order to satisfy (\ref{210}). This condition is what 
should be intuitively expected. Indeed,
when a system displays the tendency towards amorphisation, it behaves in a
"frustrated" way in the sense that the two main contributions to the
probabilities of {\em doublets} act in the opposite directions. Whenever the
modifier raises the coordination number ($m^{\prime}> m$), thus creating
more degeneracy of the given energies $E_i$ (i.e. much more possiblities of
linking the two entities $A$ and $B$). and increasing the probability of
agglomeration, the corresponding Boltzmann factor $e^{-E_2/k_BT}$ is smaller
than for the non-modified atoms, $e^{-E_1/k_BT}$, reducing the probability
of agglomeration, and vice versa.

Recalling that the stable solution corresponding to the glass-forming
tendency defines an implicit function, $T_g(p_{am})$, via the relation 
\begin{equation}  \label{214}
m(1-p_{am})(m^{\prime}e^{-E_2/k_BT_g}- me^{-E_1/k_BT_g}) -
mm^{\prime}p_{am}\,e^{-E_2/k_BT_g})\ =\ \Phi(p_{am}, T_g) = 0
\end{equation}
we can evaluate the derivative of $T$ with respect to the reduced
concentration $R={\frac {x}{1-x}}$: 
\begin{equation}  \label{215}
\frac{d T_g}{d R}\ =\ {\frac {q}{r}}\biggl[\biggl( \frac{\partial \Phi}{%
\partial p_{am}}\biggr)/ \biggl(\frac{\partial \Phi}{\partial T_g}\biggr) %
\biggr]_{\Phi(x,T_g)=0}
\end{equation}

In the limit $R = 0$ ($x=0$) the result has a particularly simple form: 
\begin{equation}  \label{216}
\biggl[ \frac{d T_g}{d R} \biggr]_{R=0}\ =\ {\frac {q}{r}}\biggl[\frac{%
dp_{am}}{dT_g} \biggl]^{-1}_{T_g=T_0} \, = {\frac {q}{r}}\biggl[ \frac{%
(2\,(m^{\prime}/m)\,e^{\frac{E_1-E_2}{kT_g}} - 1)^2} {\frac{E_2-E_1}{kT_g^2}}
\biggr]_{T_g=T_0}
\end{equation}
Inserting condition (\ref{213}), we obtain the first slope equation which is a
general relation for binary glasses $A_rX_q+R\ M_qX$ and which can be
regarded as a universal law: 
\begin{equation}  \label{217}
\biggl[\frac{dT_g}{dR} \biggr]_{R=0}\ =\ {\frac {q}{r}}{\frac{T_0}{ln\biggl[%
\frac{m^{\prime}}{m}\biggr]}}
\end{equation}
For network glasses $A_xB_{1-x}$, the formula has been obtained in \cite{l1}: 
\begin{equation}  \label{218}
\biggl[\frac{dT_g}{dx} \biggr]_{x=0}\ =\ {\frac{T_0}{ln\biggl[\frac{%
m^{\prime}}{m}\biggr]}}
\end{equation}
Equations (\ref{217}) and (\ref{218}) give the mathematical transcription of
the well-known rule mentioned at the beginning of this article \cite{r21}. 
The glass transition temperature $%
T_g$ increases with the addition of a modifier that increases the local
coordination number ($m^{\prime}> m$) [e.g. $B_2O_3$ based glasses \cite{r22}
or $Ge_xSe_{1-x}$ \cite{f19}], and decreases with the addition of a modifier
that decreases the local coordination number ($m^{\prime}< m$) [e.g. $SiO_2$
based glasses \cite{r24}].

\section{Application and discussion}

The equations (\ref{217}) and (\ref{218}) can be quite easily compared with
experimental data. We compare the slope at the $x=0$ ($R=0$) origin to a set
of experimental data among which the $T_{0}$ value and a $T_{g}$ measurement
for the lowest possible concentration $x$ (or $R$), in order to produce
approximate linearity, to be compared with the constant slope of (\ref{217})
and (\ref{218}). Therefore, we have tried to find, whenever possible,
reported glass transition temperatures of glass-forming systems which were
composed of a very high fraction of ''{\em regular}'' local configurations $%
A $ (e.g. $Q^{(4)}$ species in $SiO_{2}-Li_{2}O$ or $Se$ atoms in $%
Ge_{x}Se_{1-x}$). The lowest possible concentration is close to $5\%$ in
most of the situations, so that we can roughly approach the limit value $%
x\rightarrow 0$ of formula (\ref{217}) and (\ref{218}). From the $T_g$ and $%
T_0$ values, we can compute an `` {\em experimental value} '' of $%
m^{\prime}/m$ and compare it to the theoretical one, deduced from purely
structural considerations. In Reference \cite{r25}, R. Kerner has
demonstrated a good agreement of a close relationship with the behavior of $%
dT_{g}/dx$ at $x=0$ for the alkali-borate glass $(1-x)B_{2}O_{3}-xLi_{2}O$
and the silicate glass $(1-x)SiO_{2}-xCaO$. He predicted $m^{\prime
}/m\simeq 3/2$ in the first case, which leads to a positive derivative, and $%
m^{\prime }/m\simeq 1/3$ giving a negative one in the second one.
Nevertheless, the formula was incomplete because the factor $q/r$ was
missing, but the obtained values were correct since $q/r=1$ in these
investigated systems.

\subsection{Single-bonded systems}

\subsubsection{Chalcogenide network glasses}

Various experimental probes for equation (\ref{218}) can be obtained in very
simple simple glass forming systems, namely chalcogenide network glasses,
for which numerous experimental data are available in the literature, and $%
m=2$. For completeness, we report some values which have already been
presented elsewhere \cite{l1}. We have extended the list of investigated
systems in order to prove beyond any doubt that the impressive agreement of (%
\ref{218}) with experimental data is not a matter of coincidence. Within the
ranges (close to $x=0$, whenever possible) found in various references, the formula agrees very
well with the experimental data for the sulphur, tellurium and selenium
based glasses (Table I and II).
Indeed, the computation of the slope of $T_{g}$ at $x=0$
obtained from the measured glass transition temperatures, leads to
experimental values of $m^{\prime }/m$, which in turn can be compared to the
predicted ones, e.g. $m^{\prime }/m=2$ in $Ge_{x}Se_{1-x}$ glasses or $%
m^{\prime }/m=3/2$ in $As_{x}S_{1-x}$ systems. The structural change induced
by the growing proportion of a modifier atom is here obvious. At $x=0$, the
two-valenced atoms ($m=2$) of the $VI^{th}$ group form a network of chains
with various length; this is particularly verified in vitreous selenium. For
very low $x$ concentration, the atoms of modifier (Ge, As, Si,...) produce
cross-linking between the chains as suggested by Boolchand and Varshneya 
\cite{r35}-\cite{r36}, thus creating a new stable structural unit with
coordination number $m^{\prime }=4$ for germanium or silicon atoms (fig. 4),
or $m^{\prime }=3$ for arsenic.
\par
For the computation of the rates $m^{\prime }/m$, we have used standard $%
T_{0}$ values, found in the literature, or averaged over a set of reported
glass transition temperatures, so $T_{0}=316\ K$ for selenium \cite{r21}, $%
T_{0}=245\ K$ \cite{r21} for sulfur and $T_{0}\simeq 343\ K$ for tellurium 
\cite{r27}. As mentioned above, the sign of the derivative of $T_{g}$ with
respect to $x$ depends on the sign of $ln(m^{\prime }/m)$. In all reported
chalcogenide glasses, the rate $m^{\prime }/m$ is greater than one, the
systems display therefore an increase of the glass transition temperature
with increasing $x$, starting from $T_{0}$. Different systems are
represented in figure 5 and show that the linear approximation of (\ref{218}%
) can give a correct estimation of $T_{g}$ up to $x\simeq 0.4$ in certain
glass formers, e.g. $As_{x}S_{1-x}$, whereas the substantial increase of $%
T_{g}$ in $Ge_{x}Se_{1-x}$ systems yields a satisfying description only for $%
x\leq 0.15$. 
\par
A certain type of problem arises in the systems for which the glass-forming
region can not be extended towards $x=0$ (existence of a minimal value of $x$
for the glass-forming region); it is obvious that our formula can not be
applied when glass can not be formed in the limit $x=0$. Nevertheless, the
formula can sometimes be extrapolated down to $x=0$, when the variation of $%
T_{g}$ versus $x$ is linear for greater values of $x$. For example, the $%
(m^{\prime }/m)$ value shown in Table I for $Ga_{x}Te_{1-x}$ glasses has
been obtained by this method, because $T_{g}$ exhibit linearity over a wide
range of $x$ and can be extrapolated linearly down to $T_{0}$ \cite{r28}.
The constant slope allows then a comparison with (\ref{218}).

\subsubsection{Binary glasses}

As pointed out in ref. \cite{r25}, the slope equation (\ref{217}) seems to
be also verified in binary glass-formers. The most common single-bonded
systems are the oxide binary glasses which use the typical network formers
such as $SiO_2 $, $GeO_2$, $P_2O_5$ or $B_2O_3$. The structural change which
occurs when adding a modifier is well understood in terms of the modified
random network concept \cite{r41}. It suggest that when alkali oxide is
introduced into the glass former, the network is depolymerized through the
formation of sites bearing non-bridging oxygens (NBO). Indeed, each molecule
of the modifier $M_2O$ [$M=Li,Na,K,...$] creates in the network two ionic $%
O^\ominus M^\oplus$ sites, thus converting the covalent or partly covalent
network into a size-decreasing structure. For large amounts of modifier, the
structure reduces generally to isolated ionic species [e.g. $SiO^{4\ominus}$
units, the amorphous analogue of orthosilicates \cite{r9}].

\begin{itemize}
\item  {\em $SiO_{2}$ based glasses}
\end{itemize}

In silica based glasses, the creation of $Q^{(3)}$ species with three
covalent bridges ($m^{\prime}=3$) is proposed when adding a modifier.
Therefore $m^{\prime}/m=0.75$ in these systems. Table III show a very good
agreement of (\ref{217}) with the experimental data for various types of
modifier.
\par
Since this glass can form continously down to $x=0$ ($R=0$), it is possible
to find $T_{g}$ measurements for very low concentrations, yielding a better
agreement with the predicted value of $m^{\prime }/m$ reported in the table.
On can also remark that the dramatic decrease of the glass transition
temperature when adding a very few modifier can be mathematically explained
by the presence of the factor $T_{0}$ in the slope equation (fig. 6). 
\par
The higher the initial glass transition temperature $T_{0}$ of the network
former, the more pronounced will be the decrease of $T_{g}$. Of course,
silica based glasses, which has one of the highest $T_0$ value among glass
materials, show very well this caracteristic feature. The
fabrication of glass by ancient Egyptians is due to this fact. With
the heating techniques of that time, it was impossible to form glass from
the desert sand, made almost of $SiO_{2}$. Nevertheless, with the addition
of $10$ to $20\%$ of $K_{2}O$, obtained from the ashes of burned algae, they
could produce it quite easily, because of the sharp decrease of the glass
transition temperature.

\begin{itemize}
\item  {\em $GeO_{2}$ based glasses}
\end{itemize}

Another system behaves very similarly to the silica based glass, namely $%
GeO_2-M_2O$ systems. This system has been extensively studied because of its
unique physical properties, among which the so-called ``{\em density anomaly}%
''. Ivanov and Estropiev first reported \cite{r48} that the density of these
glasses increase with addition of alkali oxide and further studies showed
that density, as well as refractive index reach a maximum around $15-16\%$
added $Na_2O$ and then decrease \cite{r49}-\cite{r51}. Numerous structural
studies have been carried out in order to elucidate the reason of this
anomaly, among which investigations who infered the presence of $GeO_6$
octahedra within the network \cite{r52} in order to explain the density
maximum, although recent EXAFS studies \cite{r53} have clearly shown that
pressure-induced coordination changes in $v-GeO_2$ are reversable and that $%
Ge(6)$ should not observed at room pressure. Other authors believe that the
anomaly should result from an alternative structural reorganization \cite
{r54}. Micro-Raman experiments have been performed and results have been
obtained in this sense: the increase of the density with a low amount
of modifier can be related to the existence of rings of particular
sizes, namely 4- and 3-membered rings, the growing proportion of the latter
one being responsible of the density anomaly \cite{r55}.
\par
From the data which are available, we can obtain the $m^{\prime}/m$ rate
which is very close to $0.75$ in most of the systems (Table IV).
This suggests that for a very small concentration of modifier $M_{2}O$, the
coordination number of the basic $GeO_{4/2}$ tetrahedra changes into $%
m^{\prime }=3$, as in the silicate glass (fig. 7a). But in contrast with
this latter glass system, the proportion of $Q^{(3)}$ structures is not
increasing any more when $x$ is growing, as shown indirectly in fig. 8.
Indeed, the glass transition temperature shows a minimum around $x=0.02$ and
then increases. The positive derivative of $T_{g}$ versus $x$ for $x>0.02$,
means that the local coordination number of the `` {\em altered} ''
configuration is now greater than the one of the `` {\em regular} ``
configuration (supposed to be composed of a mixture of $Q^{(3)}$ and $%
Q^{(4)} $ structures), thus confirming the growing presence of $GeO_{6}$
octahedra (with $m^{\prime }=6$).

\begin{itemize}
\item  {\em $P_{2}O_{5}$ based glasses}
\end{itemize}

Other glass systems display the same agreement between the predicted value
of $m^{\prime }/m$ and the one obtained from experimental data \cite{r59}-%
\cite{r61}. This is realized in $P_{2}O_{5}$ based glasses. The available $%
T_{g}$ measurements of $(1-x)P_{2}O_{5}-xM_2O$ systems are represented in
figure 9 and compared to the equation $T_{g}=T_{0}[1+{\frac{x}{1-x}}{\frac{1%
}{ln\ 2/3}]}$. The rate of $m^{\prime }/m$ should be equal to $2/3$ in these
glass systems (fig. 7c). The basic network former $P_{2}O_{5}$ is made of
phosphor tetrahedra with four $P-O$ bonds, among which one is double bonded,
so that it is not connected to the rest of the network, therefore $m=3$ [$%
Q^{(3)}$ structure]. At the beginning, the addition of a modifier produces
the usual creation of one $NBO$, as in silicate and germanate glasses [$%
Q^{(2)}$ structure, $m^{\prime }=2$]. The rate of this structure is
increasing with the modifier concentration and reaches unity for $x=0.5$,
yielding the metaphosphate chain structure, made of corner-sharing polymeric 
$PO_{4}^{2\ominus }$ tetrahedra \cite{r62}. On this basis, we predict 
$T_g=620\ K$ for the $0.98\ P_2O_5-0.02\ Li_2O$ glass.

\begin{itemize}
\item  $B_{2}O_{3}$ based glasses
\end{itemize}

All the presented data up to now, exhibit a decrease of $T_{g}$ with growing
modifier concentration $x$, in agreement with the slope equation (\ref{217})
and the currently accepted rule which states that the increase of the
coordination number (i.e. the conectivity of the network) produces an
increase of the glass transition temperature. The well-known symmetrical
example of this rule is given by the $B_{2}O_{3}$ based glass, which
presents a positive derivative of $T_{g}$ at the origin \cite{r22}. The
addition of $M_{2}O$ [$M=Li,Na,...$] transforms the $BO_{3/2}$ triangles ($%
m=3$), which represent the basic SRO structural unit of $B_{2}O_{3}$, into $%
BO_{4}^{\ominus }$ tetrahedra ($m^{\prime }=4$, N4 species \cite{r63}, fig.
7b). The linear increase of $T_{g}$ versus the modifier concentration in
these glasses is explained by the conversion of a three-valenced network
into a four-valenced one, thus increasing the connectivity \cite{r64}. The
nature of the cation $M^{\oplus }$ seems not to have some influence for low
concentration, when $T_{g}\simeq T_{0}$ \cite{r22}. Systems with the same
concentration but with a different modifier cation, display still very close 
$T_{g}$ data (figure 10). The rate of $N4$ species is growing linearly with $%
R$ up to $R\simeq 0.5$ whatever the involved $M^{\oplus }$ cation \cite{r19}%
. For $R>0.5$, the glass tranition temperature decreases due to the growing
presence of $BO_{3/2}^{\ominus }$ triangles sharing one NBO ($m^{\prime }=2$%
) \cite{r22}. When comparing the rate $m^{\prime }/m$ obtained from
experiments with the theoretical one derived from the local structure
consideration, one obtains $m^{\prime }/m\ =\ 1.63$ instead of $4/3=1.33$. 
\par
Nevertheless, an alternative structural proposal supports the experimental
value of $1.63\simeq 5/3$. The structure of $B_{2}O_{3}$ is indeed a typical
example of well-characterized medium-range order. There is a strong
experimental evidence for the existence of larger structural groups than the
local SRO $BO_{3/2}$ triangles, namely the boroxol ring $B_{3}O_{3}$, made
of three connected $BO_{3/2}$ triangles \cite{r66}-\cite{r69}. The
spectroscopic patterns of Raman investigation, $NMR$ and neutron diffraction
exhibit for this compound sharp and well-characterized peaks, which can be
attributed either to the breathing modes of the oxygens inside the boroxol
ring (in case of Raman studies, at $808cm^{-1}$ \cite{r70,r71}) or to
the $B-O-B$ bond length inside a boroxol ring (in the case of diffraction 
\cite{r2,r68}). Whatever the proportion of this structural unit in the
network former ($0.8$ is the currently accepted value \cite{r66}-\cite{r69}), 
the coordination
number remains $m=3$. The addition of the modifier leads to the creation of
the so-called tetraborate group, even at the very beginning \cite{r63}. This
structural group is made of several three-membered rings (as the boroxol
group) sharing N4 species, with coordination number $m^{\prime }=5$. This
yields a rate of $m^{\prime }/m=1.67$ (fig. 7d).

\subsection{Relationship with the Gibbs-Di Marzio equation}

We have mentioned at the beginning of this article that Gibbs and Di Marzio
have given a formula relating $T_g$ to some structural factor, on the basis
of thermodynamical considerations. Varshneya and co-workers have 
modified this equation in order to test its validity on chalcogenide network
glasses \cite{f20,r75}. These systems satisfy all the required conditions of
Gibbs and Di Marzio's model, namely the presence of polymeric atomic chains
(as $Se$ chains in $v-Se$) which can be cross-linked by other atomic
species, such as germanium. They have expressed $T_g$ in terms of the
network average coordination number $<r>$, rather than the concentration $x$%
. $<r>$ is widely used for the description of network glasses since Phillips
has introduced this concept in his constraint theory \cite{r76}. These
authors have redefined for multicomponent chalcogenide glasses the
cross-linking density $X$ of Gibbs and Di Marzio equation (\ref{11}) as
being equal to the average coordination number of the cross-linked chain
less the coordination number of the initial chain, i.e.: $X\ =\ <r>\ -\ 2$,
and the Gibbs-Di Marzio equation can in this situation be rewritten as: 
\begin{equation}  \label{19}
T_g={\frac {T_0}{1-\beta (<r>-2)}}
\end{equation}
where $\beta$ is a system depending constant, whereas it was suggested that
the constant $\kappa$ of the initial equation (\ref{11}) is universal \cite
{f22}. Sreeram et al. fitted the constant $\beta$ to
their $T_g$ measurements by least-squares fit \cite{r75} and obtained a value which depends on
the considered system and the involved atoms.

The slope equation (\ref{218}) can be related to the Gibbs-Di Marzio
equation in the pure chalcogen limit and gives, after identification, an 
analytical expression for $\beta$.

According to Phillips \cite{r76}, one can express the average coordination
number $<r>$ in terms of the coordination number of the covalently bonded
atoms, i.e. the coordination numbers $m$ and $m^{\prime}$ of the $A$
(chalcogen atom) and $B$ configuration (modifier atom). 
\begin{equation}  \label{321}
<r>\ =\ 2(1-x)+m^{\prime}x
\end{equation}
The slope at the origin, where $x=0$ (and $<r>=2$) is then: 
\begin{equation}  \label{322}
\biggl[{\frac{d\ T_g}{d<r>}}\biggr]_{<r>=2}\ =\ {\frac {T_0}{(m^{\prime}-2)\
ln{\frac {m^{\prime}}{2}}}}
\end{equation}
In the vicinity of the pure chalcogen region ($x\rightarrow 0$), a first
order development of the Gibbs-Di Marzio equation has the following form: 
\begin{equation}  \label{324}
T_g\ \simeq\ T_0\biggl[1+\beta\ (<r>-2)\biggr]
\end{equation}
which leads by identifying (\ref{322}) and (\ref{324}) to an analytical
expression of the constant $\beta$, involving only the coordination number $%
m^{\prime}$ of the modifier atom. 
\begin{equation}  \label{325}
{\frac {1}{\beta}}\ =\ (m^{\prime}-2)\ ln({\frac {m^{\prime}}{2}})
\end{equation}

The value of $\beta$ can now be computed for different glass systems for
which the coordination number of the modifier atoms are well-known, e.g. for
chalcogenide based glasses. The possible values for $\beta$ are $0.36$ (for $%
m^{\prime}=5$), $2.47$ (for $m^{\prime}=3$) and $0.72$ (for $m^{\prime}=4$).
The latter situation corresponds to the glass $Ge_xSe_{1-x}$ and the
agreement of $\beta={\frac {1}{2\ ln2}}=0.72$ with the value obtained by a
least-squares fit of the glass transition temperatures data versus $<r>$, is
very good and close to the measurements of Varshneya and co-workers. Other
IV-VI systems behave very similarly, as seen in Table V. However, expression
(\ref{325}) is valid for binary network glasses only, but we believe that it
can be generalized for multicomponent chalcogenide network glasses,
involving at least three different types of atoms.

\subsection{Other structural contributions}

\subsubsection{Edge-sharing character}

One of the possible corrections of the equation (\ref{217}) can be produced 
by the
influence due to the edge-sharing character of the local configurations.
Binary chalcogenide glasses are the most representative systems of such a
tendency. They form indeed very easily two-membered rings (dimers) and the
fraction of dimers can be either very low (such as in $P_2S_5$ \cite{r77})
or very high as in $SiS_2$ based glasses \cite{r8}. Indeed, the proposed
long range structure in these latter systems is a chain of polymeric edge-sharing $%
SiS_{4/2}$ tetrahedra which are cross-linked by corner-sharing tetrahedra 
\cite{r78,r79}. Thus, one can consider that the local glass structure of $%
SiS_2$ and $SiSe_2$ is made of pure edge-sharing $SiX_{4/2}$ tetrahedra.
This result has been given by Tenhover on the basis of $NMR$ spectroscopy 
\cite{r78}, but also obtained by Sugai (Raman investigation and
modelization) \cite{r79b}, Vashishta and co-workers (molecular dynamics) 
\cite{r79t} and Gladden and Elliott (radial distribution function 
calculation) \cite{r3}. We have
seen that the statistical factors which appeared at the very beginning of
the construction in the expression of the probabilities of {\em doublets} (%
\ref{21})-(\ref{23}) are responsible for the presence of the term $%
ln(m^{\prime}/m)$ in (\ref{217}) and (\ref{218}). If there is an
edge-sharing tendency, the number of ways in joining together two {\em %
singlets} will be different and will modify the logarithmic expression.

The number of ways of joining by edges a {\em singlet} $A$ with coordination
number $m $ with a {\em singlet} $B$ with coordination number $m^{\prime }$
is $2\times 2\times \left( 
\begin{array}{c}
m \\ 
2
\end{array}
\right) \times \left( 
\begin{array}{c}
m^{\prime } \\ 
2
\end{array}
\right) $ in three dimensions and the probabilities can be rewritten in the
pure edge-sharing situation as: 
\begin{equation}
p_{AA2}\ =\ {\frac{2}{{\cal Z}_{2}}}\biggl[{\frac{m(m-1)}{2}}\biggr]%
^{2}(1-p)^{2}e^{-\beta E_{1}+\beta E_{e}}  \label{31}
\end{equation}
\begin{equation}
p_{AB2}\ =\ {\frac{4}{{\cal Z}_{2}}}\biggl[{\frac{m(m-1)}{2}}\biggr]\biggl[{%
\frac{m^{\prime }(m^{\prime }-1)}{2}}\biggr]p(1-p)e^{-\beta E_{2}+\beta
E_{e}}  \label{32}
\end{equation}
where $E_{e}$ is an energetical correcting factor which takes into account
the fact that the energy stored in an edge-sharing {\em doublet} is not
equal to a single bond energy $E_{i}$. 
\begin{equation}
\label{32b}
{\cal Z}_2\ =\ 2\biggl[{\frac {m(m-1)}{2}}\biggr]^2(1-p)^2e^{-\beta E_1+\beta E_e}
+4\biggl[{\frac {m(m-1)}{2}}\biggr]\biggl[{\frac {m'(m'-1)}{2}}\biggr]p(1-p)
e^{-\beta E_2+\beta E_e}
\end{equation}
The construction is
performed along the same scheme as equations (\ref{24})-(\ref{27}). The
amorphous singular solution is given by: 
\begin{equation}
p_{am}\ =\ {\frac{n(n-1)(m-1)e^{-E_{2}/kT}}{%
2n(n-1)^{2}e^{-E_{2}/kT}-m(m-1)^{2}e^{-E_{1}/kT}}}  \label{33}
\end{equation}
which exists only if: 
\begin{equation}
{\frac{m^{\prime }(m^{\prime }-1)}{m(m-1)}}\ >\ e^{(E_{2}-E_{1})/kT}
\label{33b}
\end{equation}
and the general feature of the phase diagram (fig. 5) remains the same in
this situation. The solution can be, as before, examined in the limit
condition ($x\rightarrow 0$ or $R\rightarrow 0$), where the pure A-network
exists and $p_{am}\rightarrow 0$. This yields the energy difference: 
\begin{equation}
E_{2}-E_{1}=kT_{0}\ ln\biggl[{\frac{m^{\prime }(m^{\prime }-1)}{m(m-1)}}%
\biggr]  \label{34}
\end{equation}
where $T_{0}$ is still the glass transition temperature of a pure A
configuration glass, but with pure edge-sharing local configurations. One
should note that the energetical correcting factor does not appear in (\ref
{33}) and in the forthcoming equations. The slope at the origin is then
consequently modified, but the derivation of the slope equation remains
similar to the one presented above: 
\begin{equation}
\biggl[\frac{dT}{dR}\biggr]_{R=0}\ =\ {\frac{q}{r}}{\frac{T_{0}}{ln\biggl[%
\frac{m^{\prime }(m^{\prime }-1)}{m(m-1)}\biggr]}}  \label{35}
\end{equation}
Unfortunately, there are very few experimental data at our disposal in
systems displaying a strong edge-sharing tendency, because they seem very
difficult to form for very small modifier concentrations \cite{r80}-\cite
{r82}. This is due to the strong edge-sharing tendency which is also
responsible for crystallization ease \cite{r8}.
\par
Nevertheless, some data are represented in figure 11 and they concern $%
SiSe_{2}$ and $SiS_{2}$ based glasses. As explained above, these systems
possess a high amount of dimers in the basic network former and some of the
previously cited theoretical and experimental studies propose an approximate
fraction of dimers of $53\%$, in terms of $E^{(k)}$ $NMR$ functions \cite
{r78}-\cite{r79b}($E^{(k)}$ is identified with a tetrahedron sharing $k$
common edges with its neighbors, hence $k$ runs from $0$ to $2$). The
currently accepted repartition of the $E^{(k)}$ functions for $SiX_{2}$ ($%
X=S,Se$) is: $E^{(2)}=0.29$, $E^{(1)}=0.48$, $E^{(1)}=0.23$. Figure 11
displays the available experimental data about the binary chalcogenide
systems. The different straight lines using the slope equation (\ref{35})
and corresponding to possibilities of structural modification are also
plotted. The solid line corresponds to a pure corner-sharing situation (with 
$m=4$ and $m^{\prime }=3$), whereas the dotted and dashed lines represent a
pure edge-sharing situation slope equation with respectively $m=4$ and $%
m^{\prime }=2$, and $m=4$ and $m^{\prime }=3$. Although precise information
about the glass transition temperature is lacking in the very low
modification regime (low $x$ concentration) \cite{r80}-\cite{r82}, one can
observe that a $Q^{(4)}\rightarrow Q^{(3)}$ conversion (solid line in fig. 11)
using pure
corner-sharing tetrahedra [slope equation (\ref{217}) with $m=4$ and $m'=3$] 
seems not adapted
for the description of $T_{g}$ at the origin $x=0$. The two other
possibilities seem more accurate and also support what is proposed on the
basis of $NMR$ spectroscopy by Eckert \cite{r84} and Martin \cite{r85}. In $%
SiS_{2}-Li_{2}S$ glasses, the addition of lithium sulphide produces the
conversion of $Q^{(4)}$ into $Q^{(2)}$ species (a structure with two
non-bridging sulfur atoms, i.e. a tetrahedron $SiS_{4/2}^{2\ominus }$), thus
producing a growing rate of the $Li_{2}SiS_{3}$ phases, identified with $%
Q^{(2)}$ edge-sharing dimers (the HT form of $c-Li_{2}SiS_{3}$ \cite{r84})
and $Q^{(2)}$ corner-sharing polymers (the LT form of $c-Li_{2}SiS_{3}$ \cite
{r84}). At $x=0.5$, the edge-sharing tendency remains important, the ratio
of the HT and LT phases is $1:3$ \cite{r84}. No characteristic signature of
a lithium dithiosilicate phase is observed on the $NMR$ spectroscopy patterns (the
crystalline $Li_{4}Si_{4}S_{10}$ phase), even at a concentration where this
compound should be expected (at $x\simeq 0.3$, if one refers to the oxide
analogous glass \cite{r7}). The same happens for the selenide glass. The structural
modification imposed by the presence of $Li_{2}Se$ is similar to the
sulphide system. The basic tetrahedra $Q^{(4)}$ are converted into $Q^{(2)}$
tetrahedra and a $Li_{2}SiSe_{3}$ phase occurs, made of almost $100\%$
corner-sharing $Q^{(2)}$ tetrahedra \cite{r84}.

For these two systems, the slope equation (\ref{35}) with $m=4$ and $%
m^{\prime}=2$ seems best adapted and figure 11 confirms moreless this
structural scenario.

In contrast with the lithium glass, it is possible to observe a
spectroscopic signature of a $Na_4Si_4S_{10}$ phase in the $%
(1-x)SiS_2-xNa_2S $ glasses, confirming the presence of $Q^{(3)}$ units (a $%
SiS_{4/2}^{\ominus}Na^{\oplus}$ tetrahedron), as suggested by Pradel \cite
{r86}. On this basis, a reasonable structural conversion is $%
Q^{(4)}\rightarrow Q^{(3)}$ for very low $x$ concentration (dashed line
in fig. 11). The rate of
edge-sharing structures is not decreasing when $x$ is growing and it is
still equal to $0.5$ at $x=0.5$ \cite{r82}. Therefore, one should propose for these
binary systems the pure edge-sharing slope equation with $m=4$ and $%
m^{\prime}=3$ (shaded line in fig. 11), which seems to agree with the
experimental data of $Na_2S-SiS_2$ systems. However, a glass transition
measurement for the concentration $x=0.05$, or lower, is missing, but it should
certainly be useful in order to give information about the local structural
modification. On the basis of what has been described above, we 
propose for $x=0.05$, $T_g\simeq 616\ K$ in the sodium sulfide glass
and $T_g\simeq 683\ K$ in the lithium sulfide one.

In other chalcogenide glasses, the rate of edge-sharing structures is much
lower. Typical glasses displaying a non-negligible edge-sharing tendency are
the $Sb_{2}S_{3}$ \cite{r86b}, $As_{2}S_{3}$ \cite{r87} or $GeX_{2}$ ($%
X=S,Se $) \cite{r87b} binary glasses, for which numerous experimental
measured glass transition temperatures are also available. In the case of a
mixture of corner- and edge-sharing structures, one must use the doublet
probability: 
\begin{equation}
1-p^{(1)}\ =\ {\frac{1}{2}}\biggl[2(p_{AA1}+p_{AA2})+p_{AB1}+p_{AB2}\biggr]
\label{36}
\end{equation}
where $p_{AA1}$, $p_{AA2}$, $p_{AB1}$ and $p_{AB2}$ are the probabilities of 
{\em doublets} which have been defined above. The slope equation is obtained
as previously: 
\begin{equation}
\biggl[{\frac{dT_{g}}{dR}}\biggr]_{R=0}\ =\ {\frac{q}{r}}{\frac{T_{0}}{ln%
\biggl[{\frac{m^{\prime }}{m}}\biggr]+ln\biggl[ {\frac{2+\lambda
(m-1)(m^{\prime }-1)}{2+\lambda (m-1)^{2}}}\biggr]}}  \label{38}
\end{equation}
where $\lambda =e^{-E_{e}/k_{B}T_{0}}$ uses the energetical correcting
factor $E_{e}$, if one assumes that the rate of corner- and edge-sharing
structures remains roughly constant in the low-modified r\'{e}gime. $\lambda 
$ can be related to the rate of edge-sharing {\em doublets}: 
\begin{equation}
\eta \ =\ {\frac{p_{AA2}}{p_{AA1}+p_{AA2}}}\ =\ {\frac{(m-1)^{2}\lambda }{%
2+(m-1)^{2}\lambda }}  \label{39}
\end{equation}
Figure 12 shows a typical example of such intermediate systems and gives
information about the rate of edge-sharing structures (dimers) in the
chalcogenide binary glasses $(1-x)As_{2}S_{3}-xTl_2S$. 
In the $As_2S_3$ based glass, the line using the slope equation (\ref{38})
with $\eta= 0.65$ has best agreement with experimental data. The rate of dimers
should therefore be about $0.65$ in this glass, according to figure 12.
As before, we believe that a $T_g$ measurement for $xleq 0.05$ should give
more precise information and improve the estimation of $\eta$.

\subsubsection{Influence of B-B bonds. The second slope equations}

In the first consideration of section 2, the model has been constructed only
with $A-A$ and $A-B$ {\em doublet} agglomeration. It corresponds to the most
common situations where only a very few `` {\em altered} '' configurations
should be expected when the starting structure made of almost `` {\em regular%
} '' configurations $A$ is slightly modified (low $x$ concentration).
Therefore, no $B-B$ bonds were considered. The presence of these bonds at
the very beginning of the modification (i.e. the tendency of a system to
create such bonds, even when there are very few $B$ configurations) can of
course change substantially the thermal behavior of the glass (and the final 
$T_g$) and modify the slope equations. The construction presented in the
previous section corresponded to a situation where a defect (the $B$
configuration created by a modifier) was diluted insided the whole
structure, thus leading to $A-B$ and $A-A$ {\em doublets} only. This seems
adapted for the description of binary oxide glasses or network glasses. If
the oxide ion is replaced by a larger and more polarizable ion, such as the
sulfide or the selenide ion, the local environment of a configuration,
composed of electron-rich ions ($S$,$Se$) and modifier cations ($M^{\oplus}$%
) may favour the occurence of local $B-B$ bonds.

With the notations introduced in section 2, the probability of finding a
pure single-bonded $B$ {\em doublet} is: 
\begin{equation}  \label{3321}
p_{BB1}\ =\ {\frac {{m^{\prime}}^2}{{\cal Z}_1}}p^2 e^{-\beta E_3}
\end{equation}
where $E_3$ is the $B-B$ bond energy and the ${\cal Z}_1$ the new
normalizing factor: 
\begin{equation}  \label{3322}
{{\cal Z}_1}\ =\ m^2(1-p)^2e^{-\beta E_1} + 2m m^{\prime}p(1-p)e^{-\beta
E_2}+{m^{\prime}}^2p^2e^{-\beta E_3}
\end{equation}
In the pure edge-sharing situation, the probability has the following
expression: 
\begin{equation}  \label{3323}
p_{BB2}\ =\ {\frac {2}{{\cal Z}_2}}\biggl[{\frac
{m^{\prime}(m^{\prime}-1)}{2}}\biggr]^2p^2e^{-\beta E_3+\beta E_e}
\end{equation}
The stationary solution of equation (\ref{25}) is changed and depends on the
energy $E_3$. Two energetical differences are now involved: $E_3-E_2$ and $%
E_1-E_2$. 
\begin{equation}  \label{3325}
p_{am}\ =\ \frac{m m^{\prime}e^{-E_2/k_BT_g} - m^2 e^{-E_1/k_BT_g}} {2m{%
m^{\prime}}e^{-E_2/k_BT_g} - m^2e^{-E_1/k_BT_g}-{m^{\prime}}^2e^{-E_3/k_BT_g}%
}
\end{equation}
\begin{equation}  \label{3326}
p_{am}\ =\ \frac
{mm^{\prime}(m-1)(m^{\prime}-1)e^{-E_2/k_BT_g}-m^2(m-1)^2e^{-E_1/k_BT_g}}
{2mm^{\prime}(m-1)(m^{\prime}-1)e^{-E_2/k_BT_g}-m^2(m-1)^2e^{-E_1/k_BT_g}-{%
m^{\prime}}^2(m^{\prime}-1)^2e^{-E_3/k_BT_g}}
\end{equation}
Equation (\ref{3325}) corresponds to the solution of (\ref{25}) in a pure
corner-sharing situation, whereas (\ref{3326}) is the solution of the pure
edge-sharing situation. The glass formation occurs only if $p_{am}$ is
attractive, which happens when two conditions are satisfied, namely: 
\begin{equation}  \label{3327}
{\frac {m^{\prime}}{m}}\ >\ e^\alpha\ \ \ \ \ \ \ {\frac {m}{m^{\prime}}}\
>\ e^\beta
\end{equation}
or, in the pure edge-sharing situation: 
\begin{equation}  \label{3328}
{\frac {m^{\prime}(m^{\prime}-1)}{m(m-1)}}\ >\ e^\alpha\ \ \ \ \ \ \ {\frac
{m(m-1)}{m^{\prime}(m^{\prime}-1)}}\ >\ e^\beta
\end{equation}
with $\alpha =(E_2-E_1)/k_BT$ and $\beta=(E_2-E_3)/k_BT$. When $B-B$ {\em %
doublets} are involved, $p=1$ can be attractive when the second inequation
of (\ref{3327}) or (\ref{3328}) is not satisfied. In the limit of $%
p_{am}\rightarrow 1$ (corresponding to the pure $B$ configuration network),
it is possible to obtain an expression for $E_2-E_3$, similarly to (\ref{213}%
). 
\begin{equation}  \label{3329}
E_2 - E_3 = k_B\ T_1\ ln\biggl[\frac{m}{m^{\prime}}\biggr]
\end{equation}
or: 
\begin{equation}  \label{3329b}
E_2-E_3=k_B\ T_1\ ln\biggl[{\frac {m(m-1)}{m^{\prime}(m^{\prime}-1)}}\biggr]
\end{equation}
where $T_1$ is the glass transition temperature of the pure $B$ glass
network [e.g. when $x=0.33$, the network of $(1-x)SiO_2-xNa_2O$ is supposed
to be composed of $Q^{(3)}$ structures only \cite{r7} and 
$T_1\simeq 445^oC$ \cite{r24}%
]. The implicit function $\Phi_2^c (p_{am},T_g)$ for the corner-sharing
situation is here: 
\begin{eqnarray}  \label{3330}
\Phi^c_2(p_{am},T_g)&=&p_{am}\biggl[2mm^{\prime}e^{-E_2/k_BT_g}-{m^{ \prime}}%
^2 e^{-E_3/k_BT_g} -m^2e^{-E_1/k_BT_g}\biggr] \\
&+&m^2e^{-E_1/k_BT_g}-mm^{\prime}e^{-E_2/k_BT_g}\ =\ 0  \nonumber
\end{eqnarray}
and for the edge-sharing situation: 
\begin{eqnarray}  \label{3331}
& &\Phi_2^e (p_{am},T_g)=p_{am}\biggl[2mm^{\prime}(m-1)(m^{%
\prime}-1)e^{-E_2/k_BT_g}- {m^{\prime}}^2(m^{\prime}-1)^2e^{-E_3/k_BT_g} \\
&-&m^2(m-1)^2e^{-E_1/k_BT_g}\biggr]+
m^2(m-1)^2e^{-E_1/k_BT_g}-mm^{\prime}(m-1)(m^{\prime}-1)e^{-E_2/k_BT_g}\ =\ 0
\nonumber
\end{eqnarray}
The application of (\ref{215}) and (\ref{216}) with the two limit conditions 
(when $p_{am}
\rightarrow 1$ and $p_{am}\rightarrow 0$) presented in (\ref{213}) and (\ref
{3329}), leads to the second set of slope equations: 
\begin{equation}  \label{3332}
\biggl[\frac{dT_g}{dR} \biggr]_{R=0}\ =\ {\frac {q}{r}}{\frac{T_0}{ln\biggl[%
\frac{m^{\prime}}{m}\biggr]}} \biggl[1-\biggl({\frac {m}{m^{\prime}}}\biggr)%
^{\frac {T_1-T_0}{T_0}}\biggr]
\end{equation}
in the pure corner-sharing situation and with (\ref{34}) and (\ref{3329b}): 
\begin{equation}  \label{3333}
\biggl[\frac{dT_g}{dR} \biggr]_{R=0}\ =\ {\frac {q}{r}}{\frac{T_0}{ln\biggl[%
\frac{m^{\prime}(m^{\prime}-1)}{m(m-1)}\biggr]}} \biggl[1-\biggl({\frac
{m(m-1)}{m^{\prime}(m^{\prime}-1)}}\biggr)^{\frac {T_1-T_0}{T_0}}\biggr]
\end{equation}
in the edge-sharing situation. In case of a mixture of corner- and
edge-sharing structures, the slope equation is: 
\begin{equation}  \label{3334}
\biggl[\frac{dT_g}{dR} \biggr]_{R=0}\ =\ {\frac {q}{r}}{\frac{T_0 \biggl[1-%
\biggl({\frac {m(2+\lambda (m-1)^2)}{m^{\prime}(2+\lambda
(m-1)(m^{\prime}-1))}}\biggr) ^{\frac {T_1-T_0}{T_0}}\biggr]}{ln\biggl[ {%
\frac {m^{\prime}}{m}}\biggr]+ln\biggl[{\frac {2+\lambda
(m-1)(m^{\prime}-1)}{2+\lambda (m-1)^2}} \biggr]}}
\end{equation}
One can note that the presence of $B-B$ bonds at the very beginning of the
modification, leads to a lowering of the slope, due to the presence of the
second term inside the bracketts. It can be also possible to obtain a
negative slope at the origin $R=0$ ($x=0$), despite an increase of the
coordination number $m^{\prime}>m$ and {\em vice-versa}. This is a rather
surprising result which contradicts the first slope equation and which is
against the empirical rule stating that $T_g$ is growing when the
connectivity of the network is growing, and {\em vice-versa}. 
Examples of such a reverse
behavior can be presented. They concern $B_2S_3$ and $P_2S_5$ based
glasses.

\begin{itemize}
\item  $B_{2}S_{3}$ based glasses
\end{itemize}

In the $(1-x)B_{2}S_{3}-xM_{2}S$ systems $M=Na,K,Rb,Cs$ the glass
transition temperature is decreasing when $x$ is growing, although these
compounds show a monotonic increase of tetrahedral boron units $N4$, in a
manner similar to that found in the oxide glasses (fig. 7b). The
coordination number of the $A$ configuration is $m=3$ ($BS_{3/2}$ triangle)
and $m^{\prime }=4$ for the $B$ configuration (tetrahedral boron).
\par
The connectivity of the network is increasing and the derivative should
therefore be positive. According to equation (\ref{3332}), the unexpected
negative slope can be explained by the presence of $B-B$ bonds in the very
low alkali limit, which yields a negative contribution inside the
bracketts of the expression (\ref{3332}). Indeed, the sulfide system
exhibits a very sharp increase of $N4$ species when $R$ is growing, much
more pronounced than in $B_{2}O_{3}$ based systems. Martin and co-workers
have studied the short-range order of these systems by $^{11}B\ NMR$
spectroscopy \cite{r89} and have shown that in the sodium glass, the rate of 
$N4$ species can be close to unity already for $x\simeq 0.2$, whereas the
same rate is always lower than $0.5$ in the oxide glass. According to their
spectroscopic investigation $N4\simeq 1.0$ when $x=0.2$, i.e. at measured
glass transition temperature $T_{1}=176^{o}C$ \cite{r90,r91}, and $%
T_{0}=583\ K$ \cite{r92}. Inserting these values in (\ref{3332}) yields a
negative slope and agrees with the experimental data of $B_{2}S_{3}-Na_{2}S$
systems in the very low modified glass, displayed in figure 13 (solid line).
For greater values of $x$, Martin suggests that equation (\ref{211}) is 
modified \cite{r89} and that $N4\simeq {\frac {\alpha}{2}}R$ with 
$\alpha =7.82$ instead of $N4\simeq R$ \cite{r19}. This implies that the right hand side of
equation (\ref{3332}) has to be multiplied by $3.91$. The slope equation
predicts then the right $T_g(x)$ behevior up to $x\simeq 0.2$

The fact that $N4-N4$ clusters can be produced even at the very beginning,
leading to a negative slope, explains why the dithioborate group, made of
two corner-sharing $N4$ species (fig. 14), is occuring very rapidly in these
systems \cite{r89}, whereas it appears only for $x\simeq 0.2$ in the oxide
glass \cite{r63}.

\begin{itemize}
\item  $P_{2}S_{5}$ based glasses
\end{itemize}

The available glass transition temperature data in these system concern only
the lithium based glasses. The local structure of $P_2S_5$, which has the
same st\oe chiometry than $P_2O_5$, is the same than in the oxide glass
(fig. 7c)). The SRO {\em singlet} of the network former is composed of a
phosphor atom with four $P-S$ bonds, one of them being double bonded [$%
Q^{(3)}$ structure]. The addition of a modifier ($Li_2S$) leads to the
creation of a $Q^{(2)}$ structure ($B$ configuration), sharing one
non-bridging sulfur atom, hence $m=3$ and $m^{\prime}=2$ \cite{r77}. The
slope at the origin ($x\simeq 0$) should therefore be negative, but this is
not observed experimentally. $T_g$ is increasing from $T_0=128^oC$
up to $217^oC$ (at $x=0.66$). For $x=0.5$, the proposed structure is the
metathiophosphate chain, made of almost $100\%$ $Q^{(2)}$ structures \cite
{r93}. The measured glass transition temperature is $T_1=177.4^oC$.
Inserting these values ($T_0$, $T_1$, $m$, $m^{\prime}$) in equation (\ref
{3332}) gives the slope at the origin, which is in very well agreement with the
experimental data of Kennedy \cite{r93}

Therefore, we can conclude that the $B-B$ {\em doublet} exists already at
the very beginning of the modification. This means that $P_{2}S_{5}-Li_{2}S$
systems display the tendency to form $Q^{(2)}-Q^{(2)}$ {\em doublets} inside
the network. This proposition has been also made by Kennedy and co-workers 
\cite{r93}. In order to explain the increase of $T_{g}$, these authors
suggest that the modifier may create a stronger ionic bonding formed between
fragments that could increase $T_{g}$ (i.e. a $B-B$ bonding, $Q^{(2)}$ being
the fragment which displays $T_{1}>T_{0}$). 

\section{Summary and conclusions}

In this article, we have shown that there is a strong evidence 
for the existence of a universal relationship between the glass
transition temperature and the local structure in low-modified glass
systems. Based on statistical and thermodynamical factors, the model gives
the slope at the origin $\biggl[{\frac {dT_g}{dx}}\biggr]_{x=0}$ for binary
glasses $(1-x)A_rX_q+xM_qX$ and network glasses $A_xB_{1-x}$. Let us recall
the following important results obtained:

\begin{enumerate}
\item  The model gives the correct value of the coordination numbers $m$ and 
$m^{\prime }$ of the involved atoms or the SRO structures for a great number
of systems, such as corner-sharing glasses (network and binary oxide
glasses), and edge-sharing glasses (binary chalcogenide glasses).

\item  The sharp decrease of the glass transition temperature for low
modifier concentration is explained by the presence of a high initial $T_{0}$
value in the expression of the slope equation. $SiO_{2}$ and $GeO_{2}$ based
glasses display such a behavior.

\item  The model yields an analytical expression of the constant $\beta $
appearing in the modified Gibbs-Di Marzio equation. The value computed with given $m$
and $m^{\prime }$ is in agreement with the one obtained by a least squares
fit from the experimental data by different authors.

\item  It explains that the decrease of $T_{g}$ despite an increase of the
connectivity of the glass network (and {\em vice-versa}) is due to the
presence of local $B-B$ {\em doublets}, a situation which is typical of $%
B_{2}S_{3}$ and $P_{2}S_{5}$ based glasses, for which the slope equations
agree with experimental glass transition temperature measurements.
\end{enumerate}

We believe that these encouraging results can be extended for any
concentration in network and binary glasses, and will explain mathematically
particular shapes of $T_g$ versus $x$ on the basis of structural
considerations. Besides these extensions, the model can be also applied to
three configurations $A$, $B$ and $C$, in order to investigate ternary
glass-forming systems. Such attempts will be presented in forthcoming
articles. 
\newpage
\vspace{0.5cm} 
\begin{tabular}{|c||c||c|c|c|c|}
\hline
\centering Compound & $\biggl({\frac {m^{\prime}}{m}}\biggr)_{th}$ & $\biggl(%
{\frac {m^{\prime}}{m}}\biggr)_{exp}$ & \multicolumn{2}{c|}{Obtained from} & 
Reference \\ 
&  &  & $x$ & $T_g [K]$ &  \\ \hline\hline
$Si_xTe_{1-x}$ & 2.0 & 2.11 & 0.10 & 389 & \cite{r26} \\ \hline
$Ge_xTe_{1-x}$ & 2.0 & 1.97 & 0.15 & 419 & \cite{r27} \\ \hline
$Ga_xTe_{1-x}$ & 1.5 & 1.45 & 0.20 & 528 & \cite{r28} \\ \hline\hline
$As_xS_{1-x}$ & 1.5 & 1.54 & 0.11 & 307 & \cite{r29} \\ \hline
$Ge_xS_{1-x}$ & 2.0 & 1.72 & 0.10 & 290 & \cite{r30} \\ \hline
\end{tabular}

\vspace{0.5cm} {\em Table I: Different tellurium and sulfur based glasses.
Comparison between the theoretical value of $m^{\prime}/m$ and the
experimental value of $m^{\prime}/m$ deduced from the slope using data of $%
T_g$ for the lowest available concentration $x$. $T_0$ has been taken as 
$343\ K$ in tellurium and $245\ K$ in sulfur.}

\vspace{0.5cm} 
\begin{tabular}{|c||c||c|c|c|c|}
\hline
\centering Compound & $\biggl( {\frac {m^{\prime}}{m}}\biggr)_{th}$ & $%
\biggl({\frac {m^{\prime}}{m}}\biggr)_{exp}$ & \multicolumn{2}{c|}{Obtained
from} & Reference \\ 
&  &  & $x$ & $T_g [K]$ &  \\ \hline\hline
$Ge_xSe_{1-x}$ & 2.0 & 2.04 & 0.05 & 336 & \cite{f19} \\ \hline
$Si_xSe_{1-x}$ & 2.0 & 2.04 & 0.05 & 336 & \cite{r31} \\ \hline
$As_xSe_{1-x}$ & 1.5 & 1.54 & 0.003 & 318 & \cite{r32} \\ \hline
$Sb_xSe_{1-x}$ & 1.5 & 1.31 & 0.15 & 493 & \cite{r33} \\ \hline
$P_xSe_{1-x}$ & 2.5 & 2.53 & 0.05 & 333 & \cite{r34} \\ \hline
\end{tabular}

\vspace{0.5cm} {\em Table II: Different selenium based glasses. Comparison
between the theoretical value of $m^{\prime}/m$ and the experimental value
deduced from the slope using data of $T_g$ for the lowest available
concentration $x$. $T_0$ has been teken as $316\ K$.}

\vspace{0.5cm}

\vspace{0.5cm} 
\begin{tabular}{|c|cccc|}
\hline
System & $Na_2O$ & $K_2O$ & $Rb_2O$ & $PbO$ \\ \hline
$\biggl({\frac {m^{\prime}}{m}}\biggr)_{exp}$ & 0.75 & 0.78 & 0.66 & 0.74 \\ 
\hline
obtained from the &  &  &  & extra- \\ 
concentration $x$ & 0.05 & 0.05 & 0.08 & polation \\ \hline
Reference & \cite{r24} & \cite{r45} & \cite{r46} & \cite{r47} \\ \hline
\end{tabular}

\vspace{0.5cm} {\em Table III: The $m^{\prime}/m$ rate deduced from
experimental data, compared to the theoretical value of $0.75$ in $SiO_2$
based glasses, with the average value of $T_0=1463\ K$} \vspace{0.5cm}

\vspace{0.5cm} 
\begin{tabular}{|c|cccc|}
\hline
System & $Li_2O$ & $K_2O$ & $Rb_2O$ & $Cs_2O$ \\ \hline
$\biggl({\frac {m^{\prime}}{m}}\biggr)_{exp}$ & 0.85 & 0.73 & 0.71 & 0.68 \\ 
\hline
obtained from the &  &  &  &  \\ 
concentration $x$ & 0.01 & 0.02 & 0.02 & 0.02 \\ \hline
Reference & \cite{r56} & \cite{r57} & \cite{r57} & \cite{r57} \\ \hline
\end{tabular}

\vspace{0.5cm} {\em Table IV: The $m^{\prime }/m$ rate deduced from
experimental data, compared to the theoretical value of $0.75$ in $GeO_{2}$
based glasses, with the average value of $T_{0}=820\ K$} \vspace{0.5cm}

\vspace{0.5cm} 
\begin{tabular}{|c|c|c|c|}
\hline
\centering System & $\beta_{fit}$ & Correlation & Reference \\ 
&  & coefficient &  \\ \hline
$Ge_xSe_{1-x}$ & 0.74 & 0.993 & \cite{f19} \\ 
$Ge_xSe_{1-x}$ & 0.72 & 0.988 & \cite{f20} \\ 
$Ge_xSe_{1-x}$ & 0.65 & 0.993 & \cite{r75} \\ 
$Ge_xS_{1-x}$ & 0.73 & 0.998 & \cite{r30} \\ 
$Si_xSe_{1-x}$ & 0.81 & 0.997 & \cite{r31} \\ 
$P_xSe_{1-x}$ & 0.138 & 0.902 & \cite{r34} \\ \hline
\end{tabular}

\vspace{0.5cm} {\em Table V: Computed values of the constant $\beta$ of the
modified Gibbs-Di Marzio equation, obtained from a least-squares fit, for
different chalcogenide glass systems. They can be compared to $\beta =0.72$ (%
$m^{\prime}=4$) or $\beta = 0.36$ ($m^{\prime}=5$). Some data are taken from
[1]} \vspace{0.5cm}
\newpage
\begin{figure}
\begin{center}
\psfig{figure=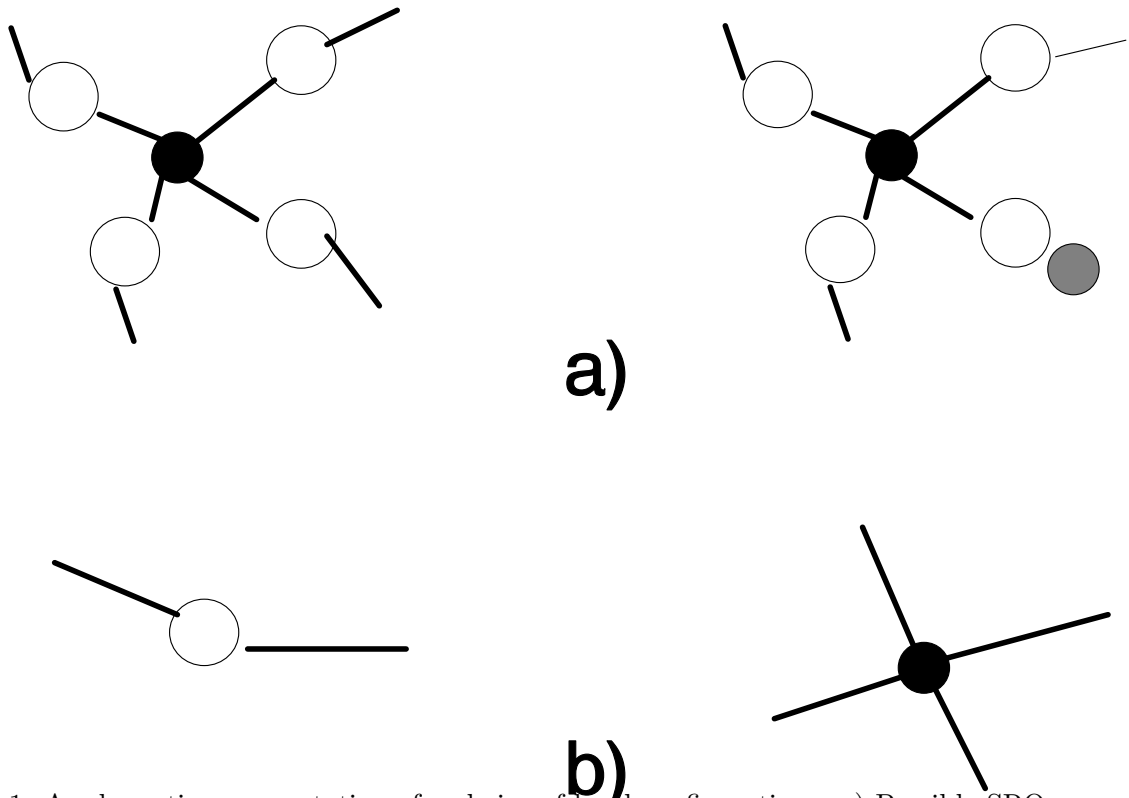,width=\linewidth}
\caption{A schematic representation of a choice of local configurations.
a) Possible SRO star-like configurations in $SiO_2-Li_2O$ glasses: a 
"{\em regular}" tetrahedron $SiO_{4/2}$ [a $Q^{(4)}$ structure] and an 
"{\em altered}" tetrahedron $SiO_{4/2}^{\ominus}$ [a $Q^{(3)}$ structure]
b) $Ge_xSe_{1-x}$ glass: a "{\em regular}" atom $Se$ and an "{\em altered}"
atoms Ge}
\end{center}
\end{figure}
\begin{figure}
\begin{center}
\psfig{figure=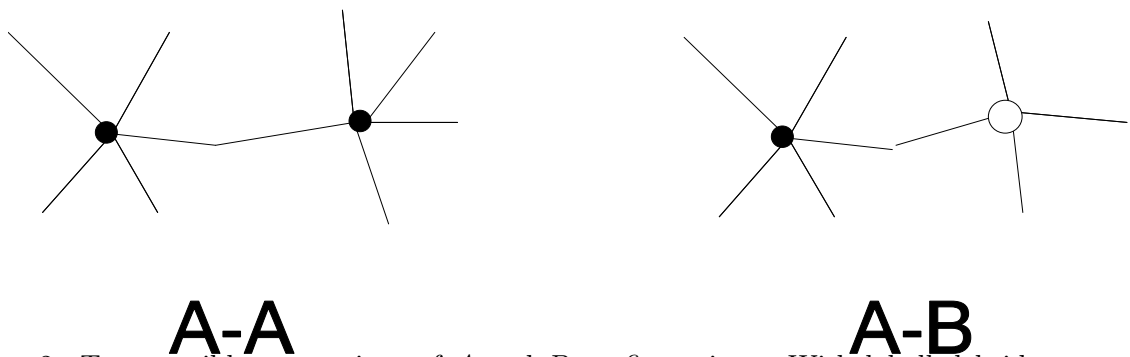,width=\linewidth}
\caption{Two possible connections of $A$ and $B$ configurations. With 
labelled bridges, there are $2\times m \times m^{\prime}$ ways to form the 
same doublet $A-B$}
\end{center}
\end{figure}
\begin{figure}
\begin{center}
\psfig{figure=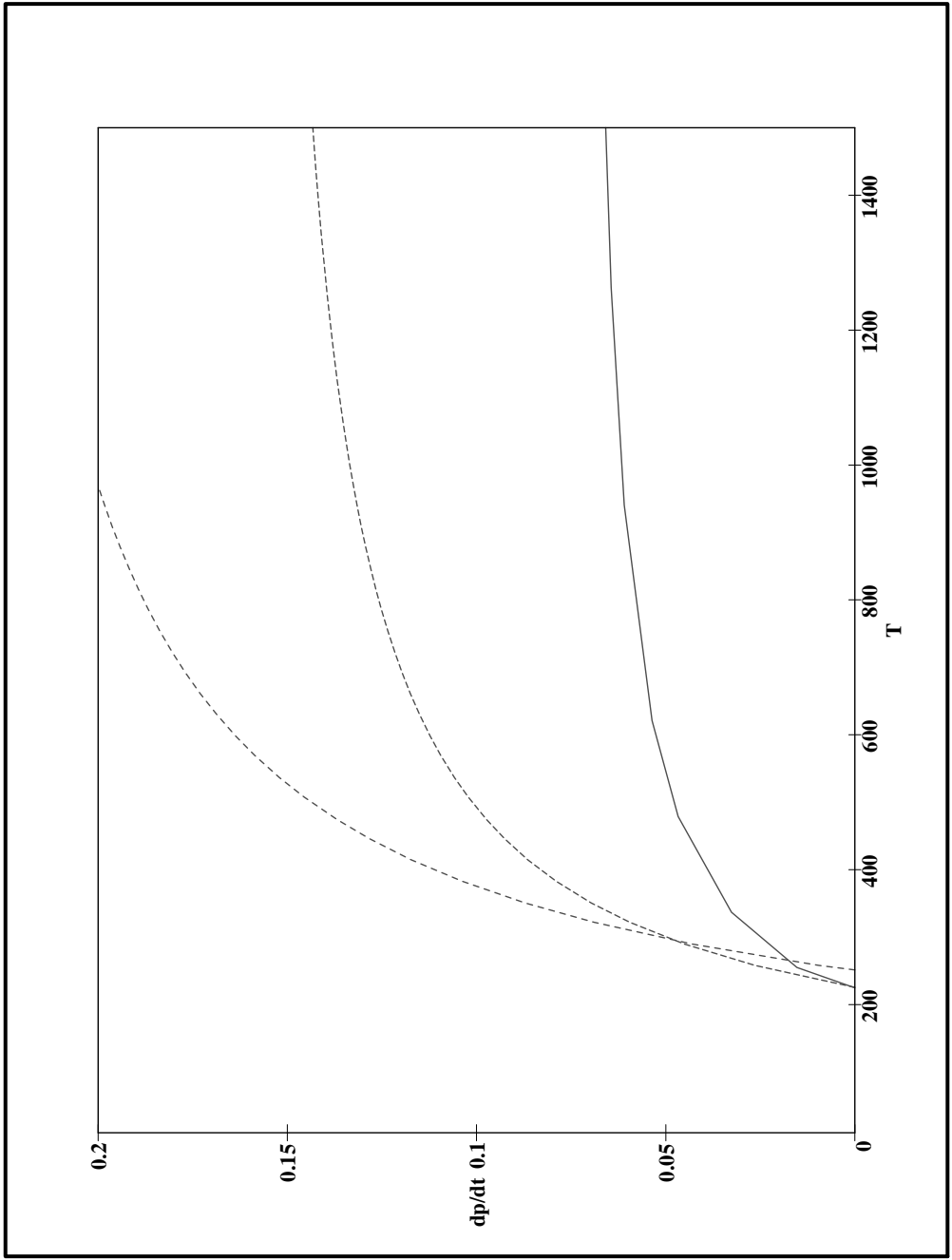,width=\linewidth}
\caption{An example of the arrest of the temporal fluctuations of 
structure. The plot represents the equation (2.5) with $m=2$, $m'=4$, 
$E_2-E_1=0.0189\ eV$ [$T_0=316\ K$, $Ge_xSe_{1-x}$ glass] and $p=0.02$ 
(solid line), $p=0.05$ (dashed line) and $p=0.10$ (dotted line)}
\end{center}
\end{figure}
\begin{figure}
\begin{center}
\psfig{figure=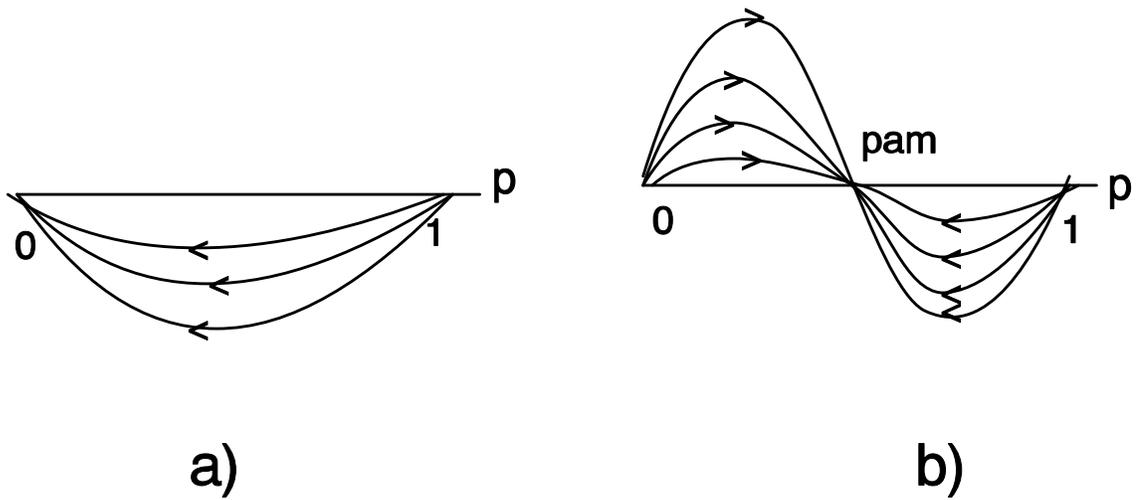,width=\linewidth}
\caption{The phase diagrams of the equ.(6): a) ${\frac {m^{\prime}}{m}<
e^\alpha}$ (phase separation) b) ${\frac
{m^{\prime}}{m}} >e^{\alpha}$ (glass formation)}
\end{center}
\end{figure}
\begin{figure}
\begin{center}
\psfig{figure=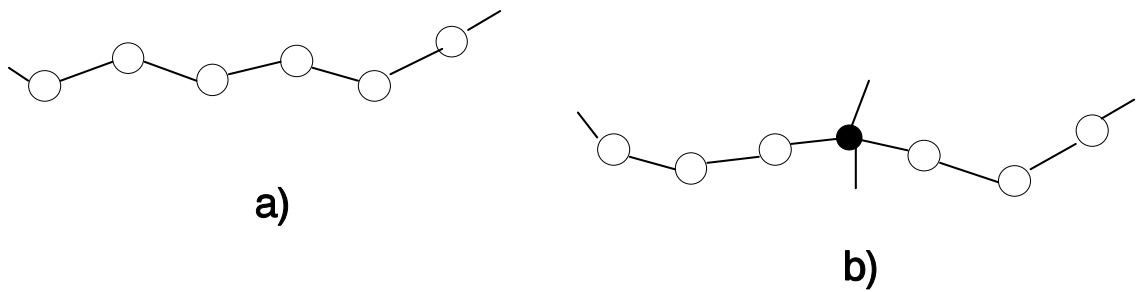,width=\linewidth}
\caption{a) A selenium chain b) a selenium chain with a cross-linking silicon 
atom}
\end{center}
\end{figure}
\begin{figure}
\begin{center}
\psfig{figure=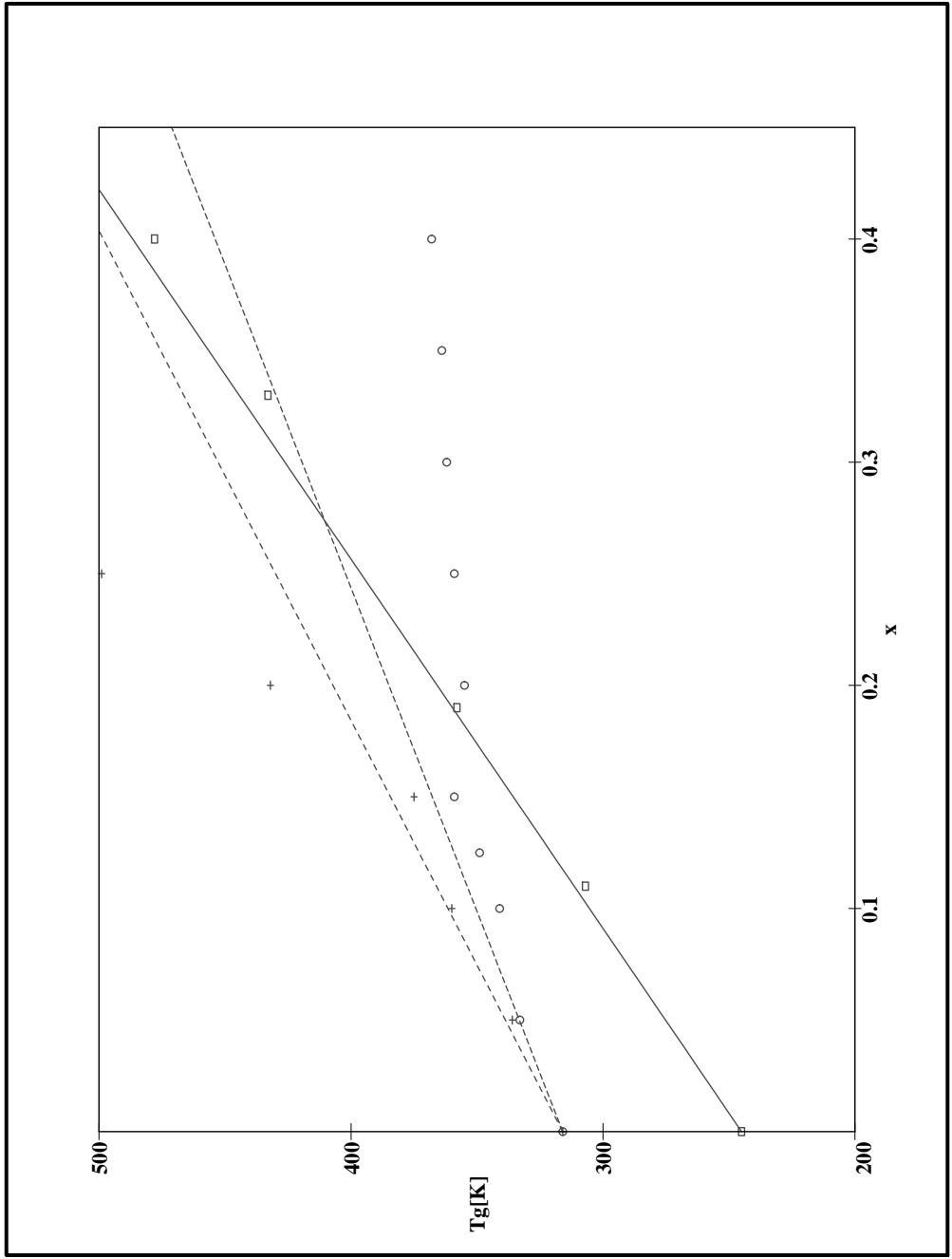,width=\linewidth}
\caption{ Glass transition temperature [in K] versus concentration $x$ for
typical glass systems, $As_xS_{1-x}$ [$\Box$], $P_xSe_{1-x}$ [$o$] and $%
Ge_xSe_{1-x}$ [$+$]. The plotted values are reported in [20], [47] and [52]. The
straight lines correspond respectively to the equations $T_g=245(1+{\frac
{x}{ln(3/2)}})$ (solid line), $T_g=316(1+{\frac {x}{ln(5/2)}})$ (dashed
line) and $T_g=316(1+{\frac {x}{ln(2)}})$ (dotted line). }
\end{center}
\end{figure}
\begin{figure}
\begin{center}
\psfig{figure=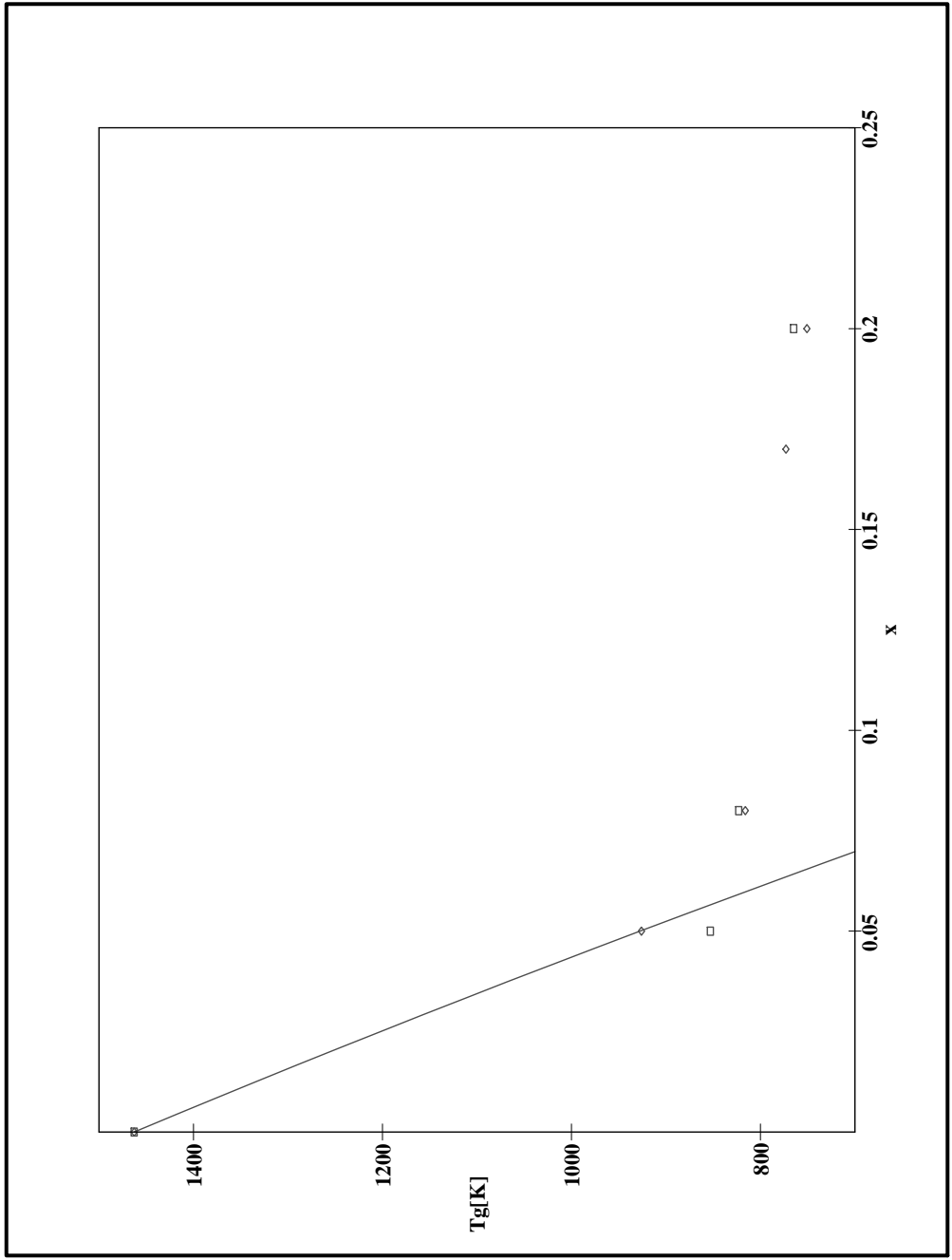,width=\linewidth}
\caption{{Glass transition temperature [in K] of $(1-x)SiO_{2}-xNa_{2}O$
[$\Diamond$] and $(1-x)SiO_{2}-xK_{2}O$ [$\Box$] systems versus the
concentration $x$. The line represents the slope equation $T_{g}=1463(1+{%
\frac{2x}{(1-x)ln(3/4)}})$ The plotted $T_{g}$ data are taken from [42]
and [56].}}
\end{center}
\end{figure}
\begin{figure}
\begin{center}
\psfig{figure=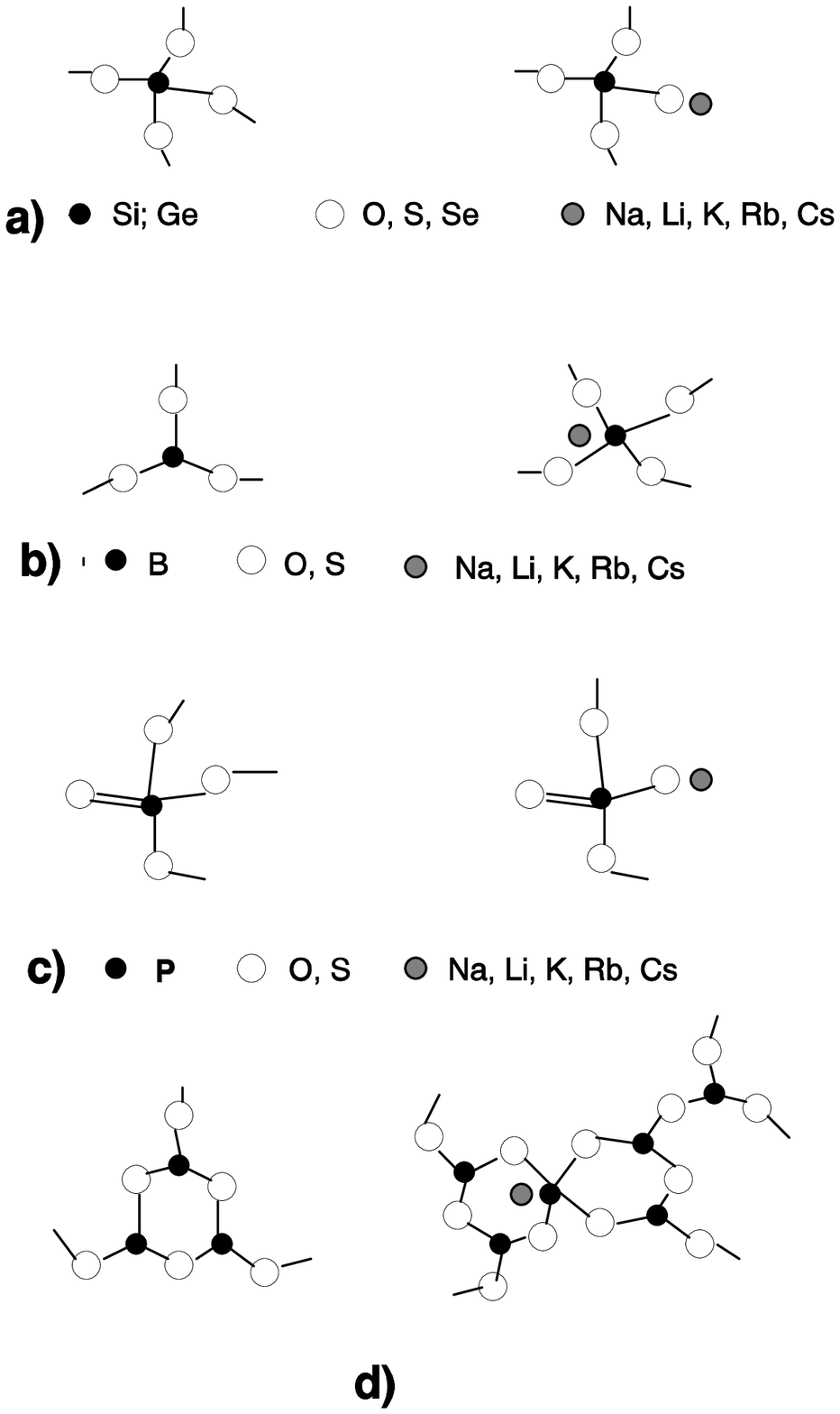,width=\linewidth}
\caption{a) IV-VI binary glasses. The
SRO structures are the $Q^{(4)}$ and $Q^{(3)}$ tetrahedra and eventually the 
$Q^{(2)}$ tetrahedra with two non-bridging atoms ($m=4$, $m^{\prime}=3$). b)
SRO structures of a boron glass. A $BX_{3/2}$ triangle ($X=O,S$) and a
four-coordinated boron (N4 unit) ($m=3$, $m^{\prime}=4$). c) SRO structure
of phosphor based glasses $P_2X_5$ ($X=O,S$) ($m=3$,$m^{\prime}=2$). d)
Possible MRO in $B_2O_3$ based glasses. The boroxol ring ($m=3$) and the
tetraborate group ($m^{\prime}=5$).}
\end{center}
\end{figure}
\begin{figure}
\begin{center}
\psfig{figure=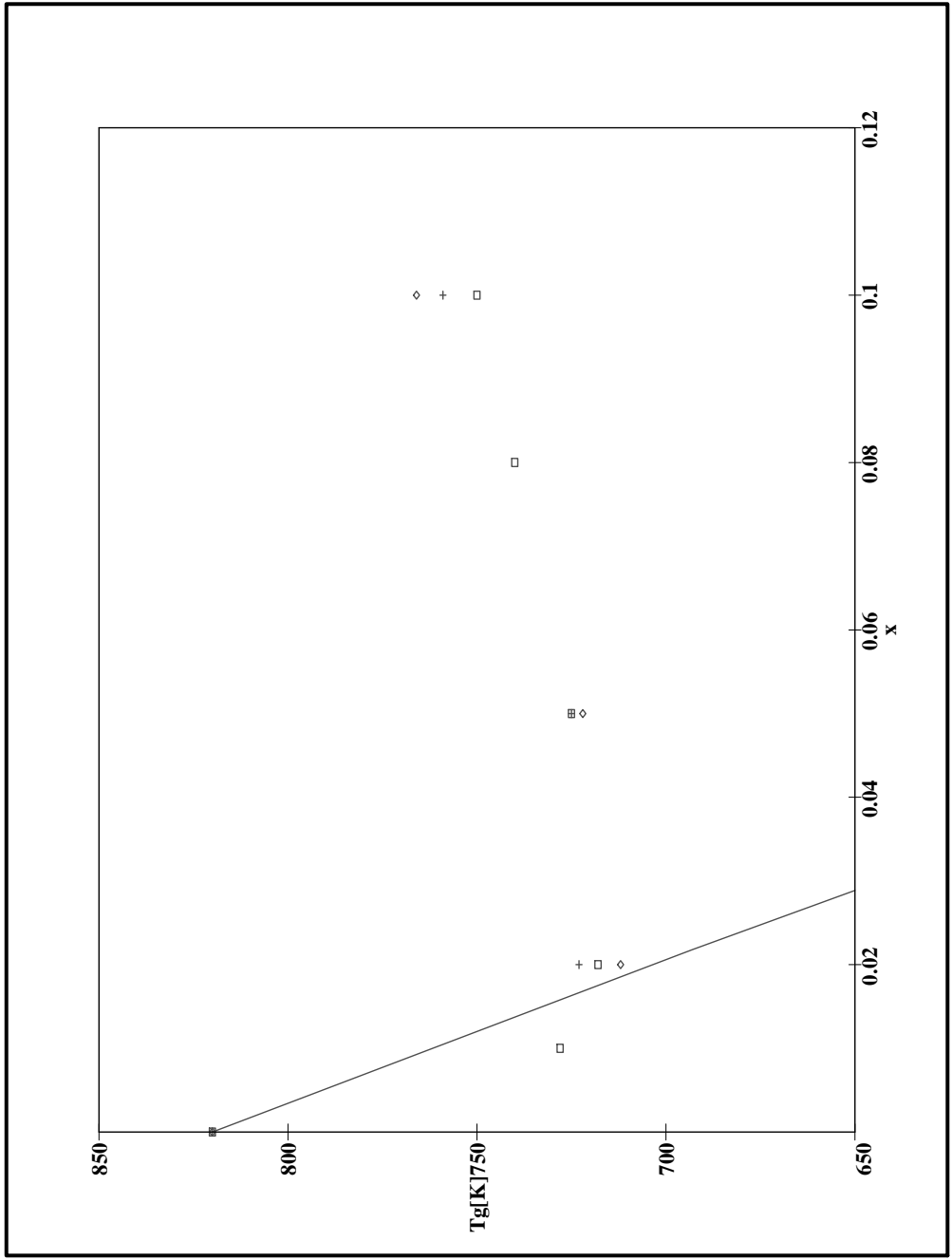,width=\linewidth}
\caption{Glass transition temperature [in K] versus the concentration $x$ in 
$(1-x)GeO_2-xM_2O$ glasses with $M=Li$ [$\Box$], $K$ [$\Diamond$] and $Rb$ [$%
+$]. The line represents the equation $T_g=820(1+{\frac {2x}{(1-x)ln(3/4)}})$%
. The data are reported in [67] and [68].}
\end{center}
\end{figure}
\begin{figure}
\begin{center}
\psfig{figure=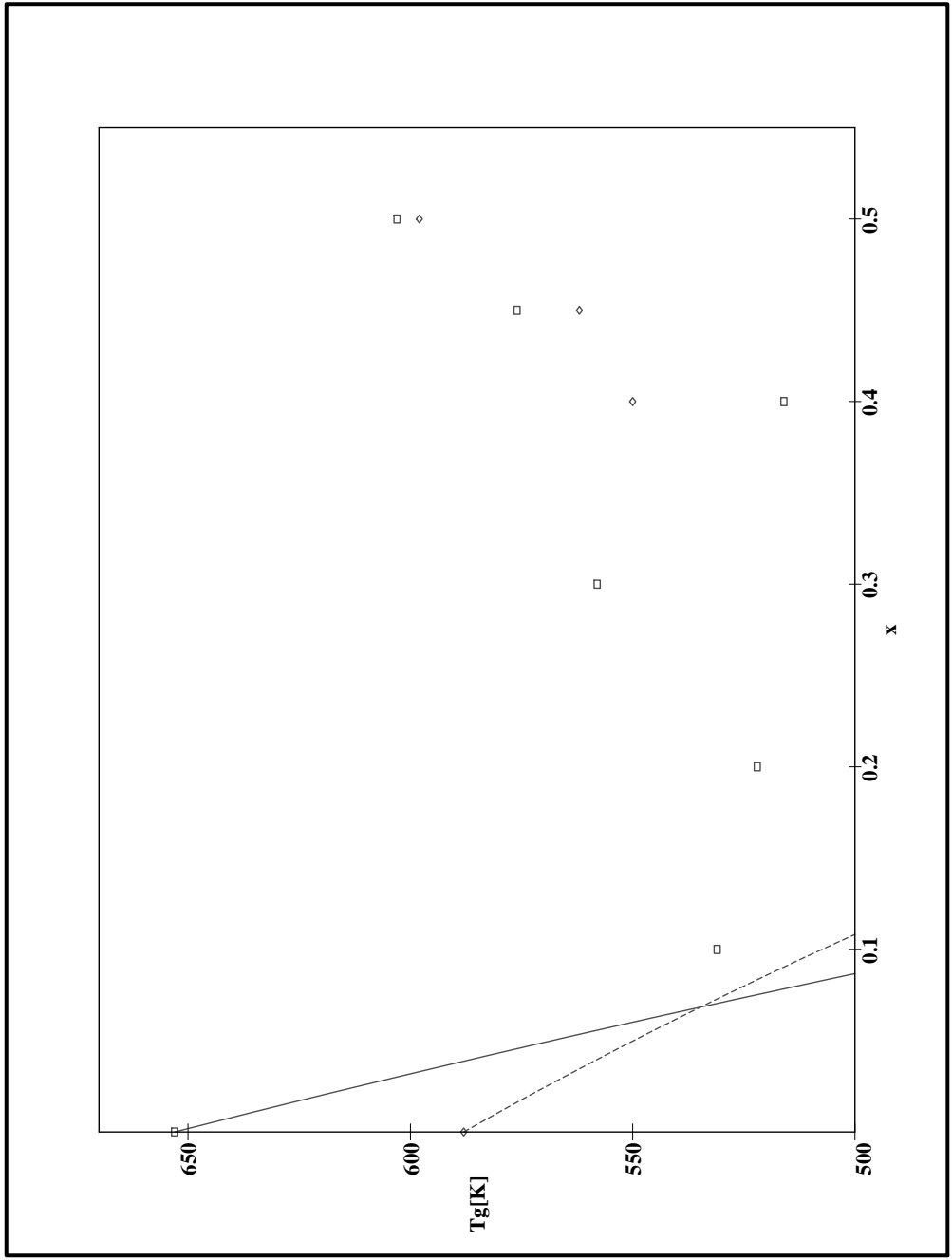,width=\linewidth}
\caption{{Glass transition temperature [in K] of $%
(1-x)P_{2}O_{5}-xM_{2}O $ glasses versus modifier concentration $x$ with $%
M=Li$ [$\Box $], $Pb_{1/2}$ [$\Diamond $]. The lines represented correspond
to $T_{g}=653(1+{\frac{x}{(1-x)ln(2/3)}})$ (solid line, $M=Li$) and to $%
T_g=588(1+{\frac {x}{2(1-x)ln(2/3) }})$ (dotted line, $M=Pb_{1/2}$). The
plotted data are given in reference [69] and [71].}}
\end{center}
\end{figure}
\begin{figure}
\begin{center}
\psfig{figure=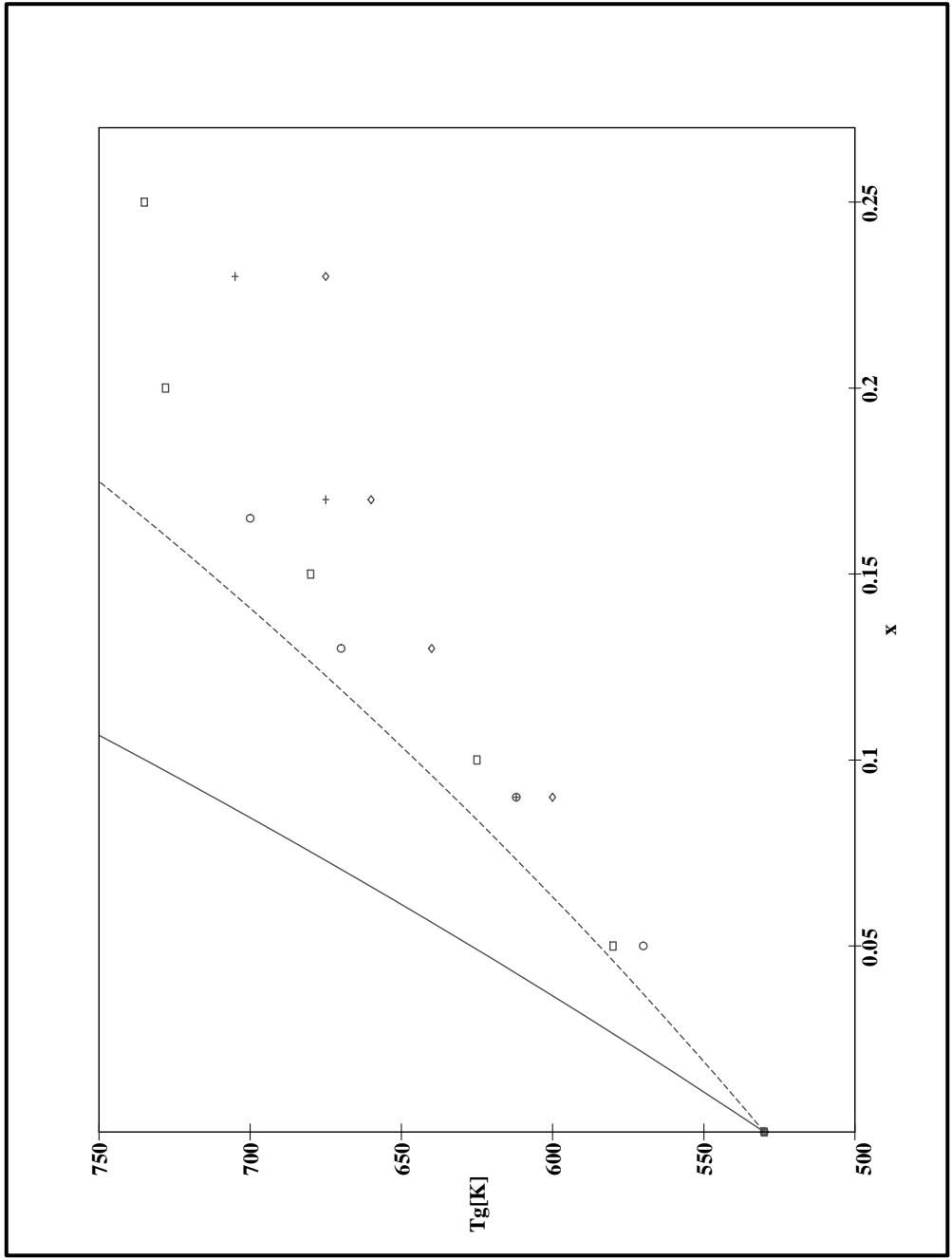,width=\linewidth}
\caption{{Glass transition temperature [in K] versus the concentration $%
x $ in $(1-x)B_{2}O_{3}-xM_{2}O$ glasses, with $M=Li$ [$o$], $Na$ [$\Box $], 
$K $ [$+$], $Rb$ [$\Diamond $] and $Cs$ [$.$]. The lines represent the
equations $T_{g}=530(1+{\frac{x}{(1-x)ln(5/3)}})$ (dotted line) and $%
T_{g}=530(1+{\frac{x}{(1-x)ln(4/3)}})$ (solid line). The plotted values are
taken from [41].}}
\end{center}
\end{figure}
\begin{figure}
\begin{center}
\psfig{figure=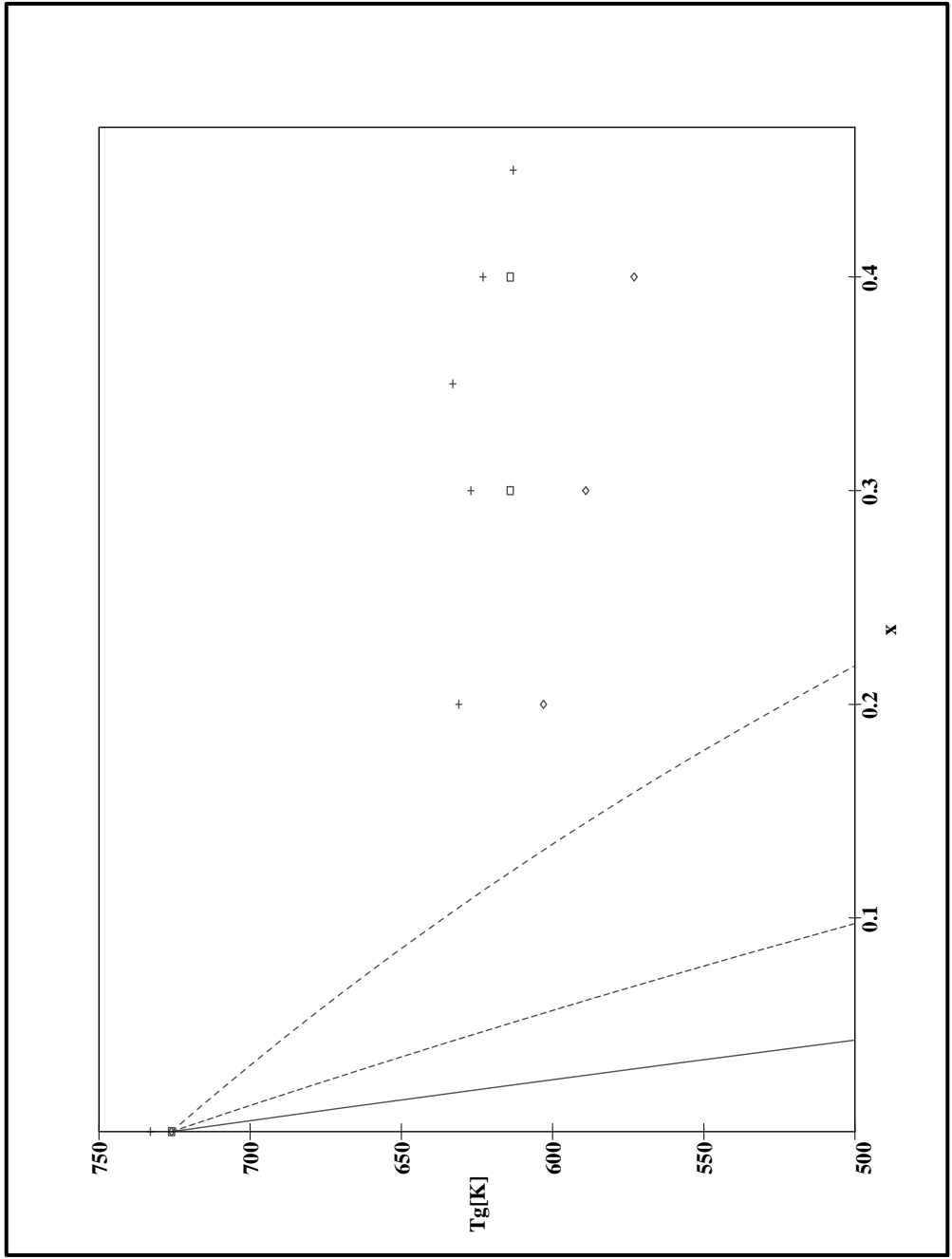,width=\linewidth}
\caption{{Glass transition temperature [in K] of $(1-x)SiS_{2}-xM_{2}S$, 
$M=Li$ [$\Box $] $M=Na$ [$\diamondsuit $] and $(1-x)SiSe_{2}-xLi_{2}Se$
glasses [$+$]. The represented lines correspond to $T_{g}=726(1+{\frac{2x}{%
(1-x)ln(3/4)}})$ (solid line, pure corner-sharing and $m^{\prime }=3$), $%
T_{g}=726(1+{\frac{2x}{(1-x)ln(2/12)}})$ (dotted line, pure edge-sharing and 
$m^{\prime }=2$) and $T_{g}=726(1+{\frac{2x}{(1-x)ln(6/12)}})$ (shaded line,
pure edge-sharing and $m^{\prime }=3$). The data are taken from [88]-[90].}}
\end{center}
\end{figure}
\begin{figure}
\begin{center}
\psfig{figure=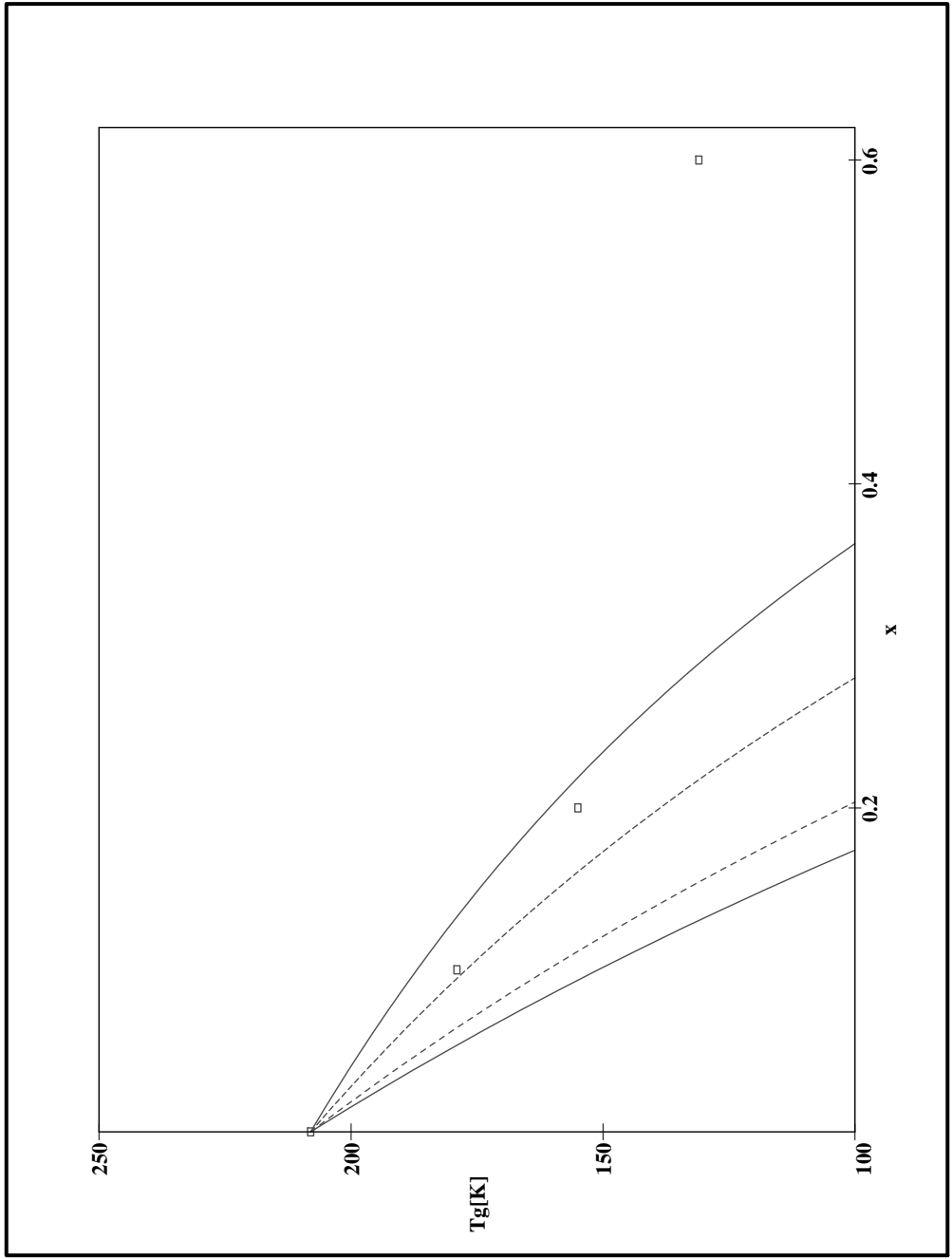,width=\linewidth}
\caption{Glass transition temperature of $(1-x)As_2S_3-xTl_2S$ glasses. The
solid lines correspond to the pure edge-and corner-sharing situations, using
the slope equation (2.16) and (3.12). Data are taken from [95]. Intermediate
situations are plotted and correspond to ($\lambda=0.1$, $\eta=0.17$, dotted
line) and ($\lambda=0.7$, $\eta=0.58$, dashed line).}
\end{center}
\end{figure}
\begin{figure}
\begin{center}
\psfig{figure=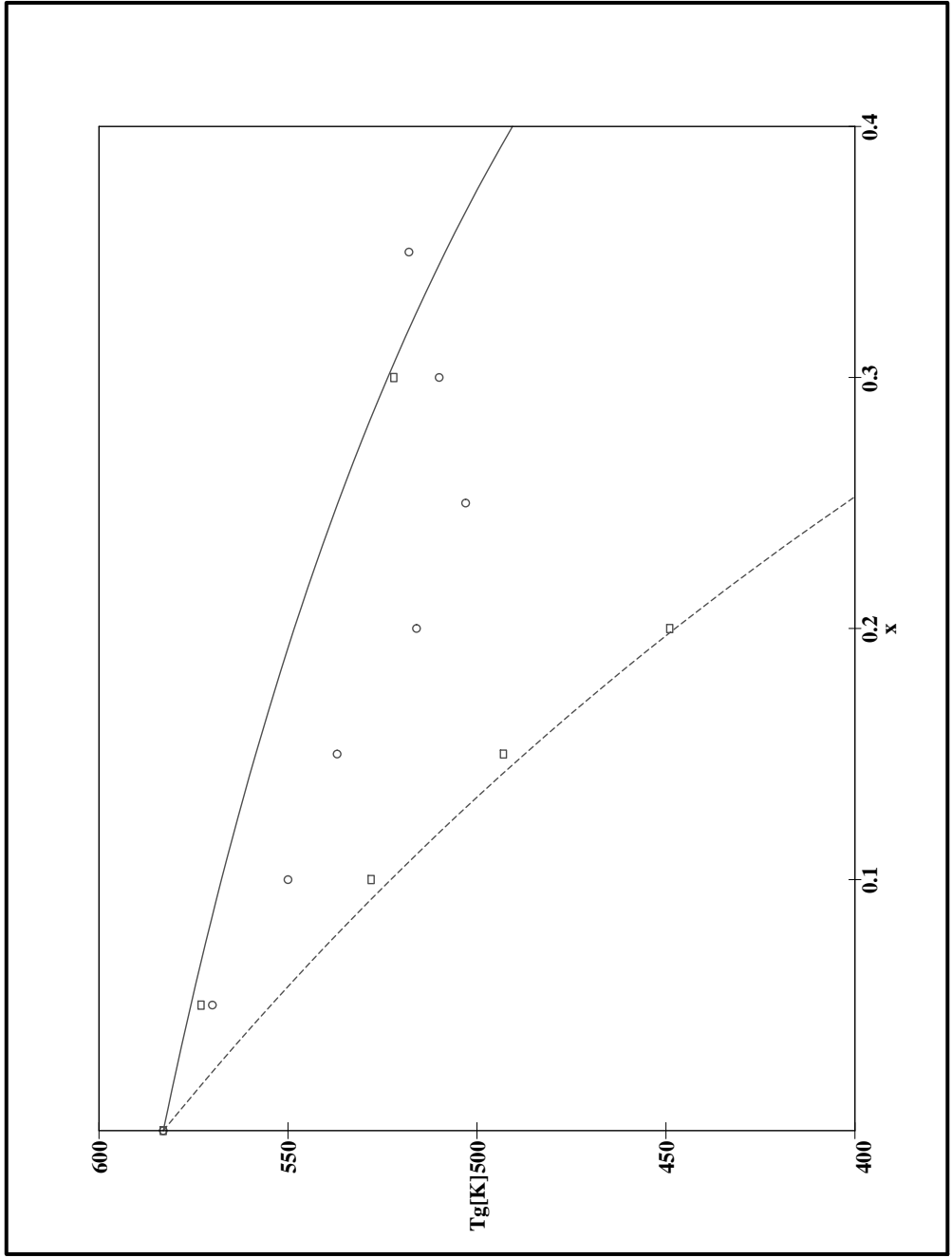,width=\linewidth}
\caption{{Glass transition temperature [in $K$] of $(1-x)B_{2}S_{3}-xM_{2}S$
glasses with $M=Na$ [$\Box $] and $K$ [$o$]. The solid line 
corresponds to the slope equation (3.27) with $m=3$, $m^{\prime }=4$, 
$T_{0}=583\ K$ and $T_{1}=449\ K$ (for sodium systems). Data are taken 
from [99]. The dotted line represents a modified slope equation using Martin's
correcting factor $\alpha =7.82$ [92] for $Na$ based glasses.}}
\end{center}
\end{figure}
\newpage
\begin{figure}
\begin{center}
\psfig{figure=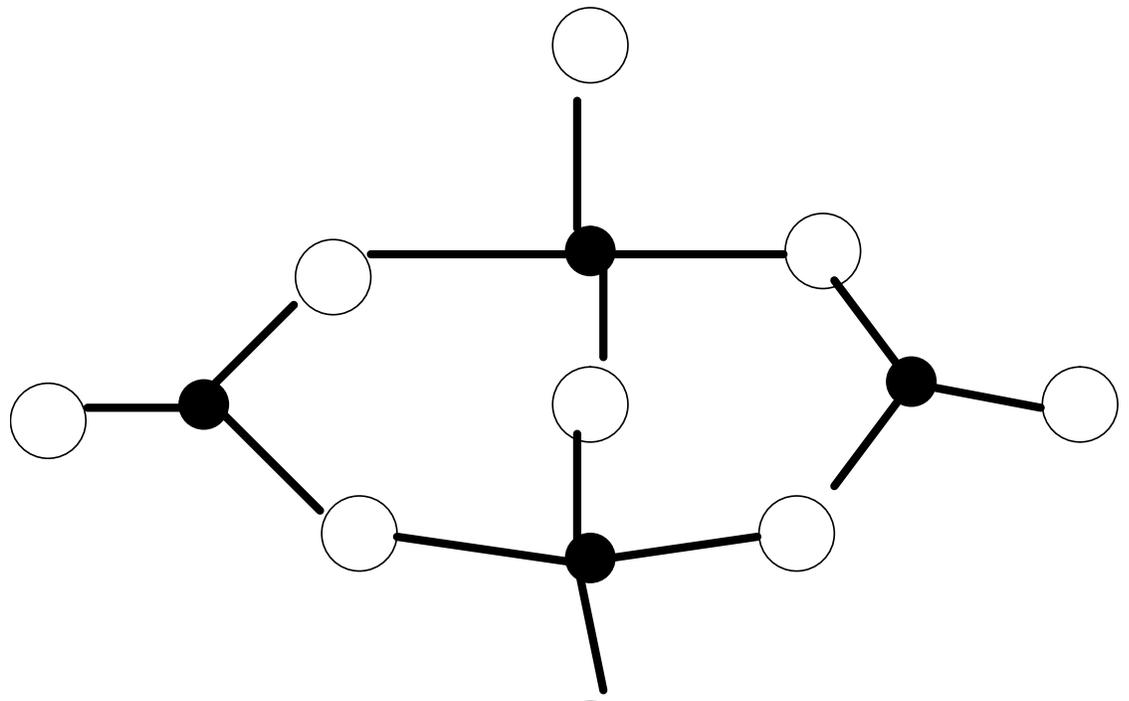,width=\linewidth}
\caption{The dithioborate group proposed
as possible MRO structure in $B_2S_3-M_2S$ glasses [92]. This structure
is made of corner-sharing $B-B$ doublets ($N4$ species).}
\end{center}
\end{figure}
\begin{figure}
\begin{center}
\psfig{figure=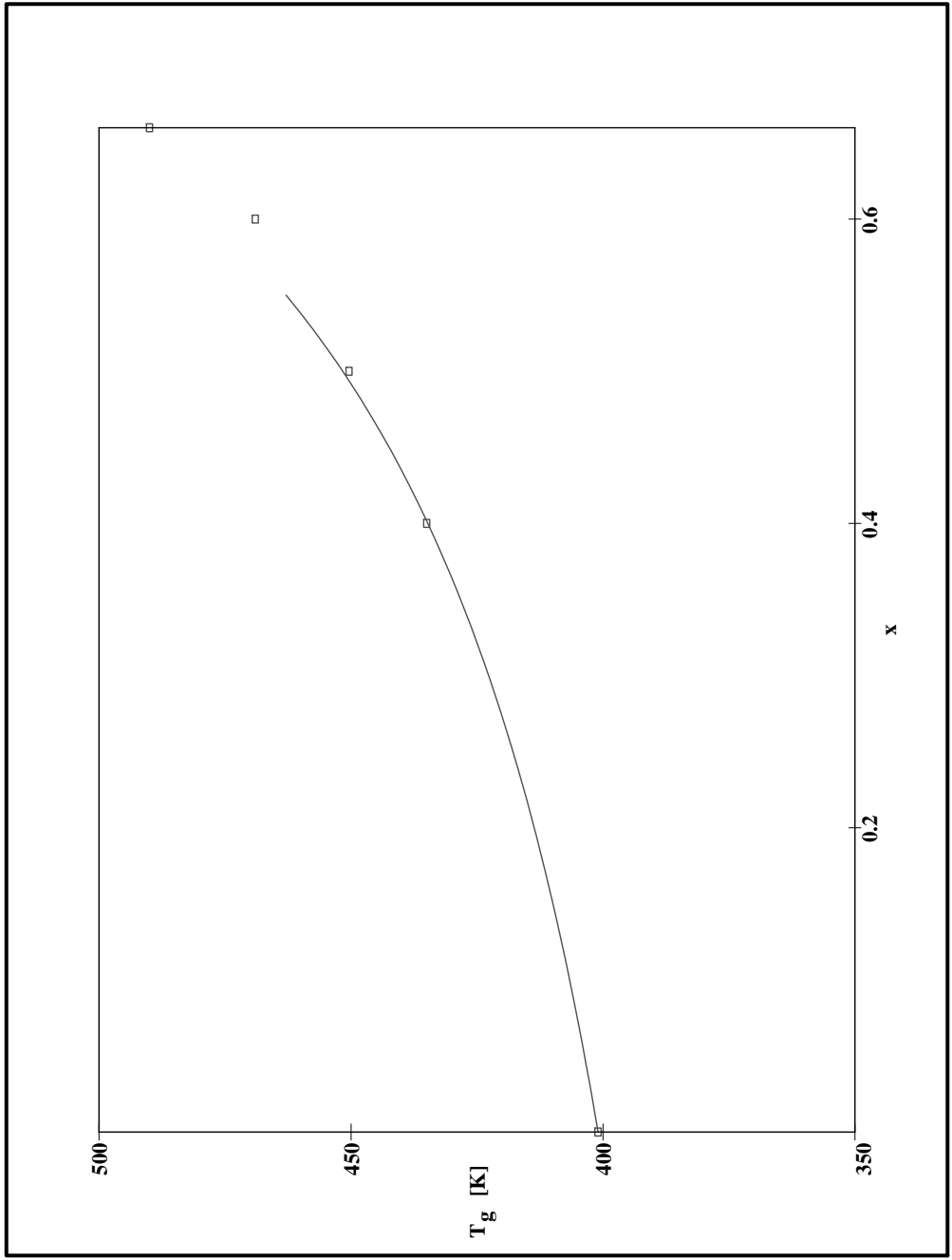,width=\linewidth}
\caption{Glass transition temperature [in K] of $(1-x)P_{2}S_{5}-xLi_{2}S$
systems. The line represents has a slope (3.27) with $m=3$, $m'=2$, $T_0=128^oC$
and $T_1=177.4^oC$. Data are reported in reference [102].}
\end{center}
\end{figure}

\end{document}